\documentclass[a4paper,conference]{IEEEtran}
\usepackage{times,amsmath,amssymb}
\usepackage{slashbox}
\usepackage{graphicx,subfigure}
\usepackage{xcolor}
\usepackage{setspace}
\usepackage[export]{adjustbox}
\usepackage{multirow}
\usepackage{bm}
\usepackage{stmaryrd}
\usepackage{soul}   
\usepackage{epstopdf}

\usepackage[ruled,vlined,linesnumbered]{algorithm2e}
\SetKwFunction{KwFn}{Fn}
\SetKwInOut{Input}{input}
\SetKwInOut{Output}{output}

\newcommand{\Dcal}{{\mathcal D}}

\newcommand{\Ical}{{\mathcal I}}

\newcommand{\Rcal}{{\mathcal R}}

\newcommand{\Tcal}{{\mathcal T}}

\newcommand{\Wcal}{{\mathcal W}}

\newtheorem{defn}{Definition}
\begin{document}

\title{Sparse Index Tracking via Topological Learning}
\author{\IEEEauthorblockN{Anubha Goel\IEEEauthorrefmark{1}\IEEEauthorrefmark{2}, Puneet Pasricha\IEEEauthorrefmark{3}\IEEEauthorrefmark{4}, Juho Kanniainen\IEEEauthorrefmark{1}}
\IEEEauthorblockA{\IEEEauthorrefmark{1}Computing Science/Financial Computing and Data Analytics Group, Tampere University, Finland\\
\IEEEauthorrefmark{2}Corresponding author. Email: anubha.goel@tuni.fi\\
\IEEEauthorrefmark{3}Ecole Polytechnique F\'ed\'erale de Lausanne, CH 1015 Lausanne, Switzerland\\
\IEEEauthorrefmark{4}Department of Mathematics, Indian Institute of Technology Ropar, Punjab-140001, India}\\

}

\maketitle

\begin{abstract}
In this research, we introduce a novel methodology for the index tracking problem with sparse portfolios by leveraging topological data analysis (TDA). Utilizing persistence homology to measure the riskiness of assets, we introduce a topological method for data-driven learning of the parameters for regularization terms. Specifically, the Vietoris–Rips filtration method is utilized to capture the intricate topological features of asset movements, providing a robust framework for portfolio tracking. Our approach has the advantage of accommodating both $\ell_1$ and $\ell_2$ penalty terms without the requirement for expensive estimation procedures. We empirically validate the performance of our methodology against state-of-the-art sparse index tracking techniques, such as Elastic-Net and SLOPE, using a dataset that covers 23 years of S\&P 500 index and its constituent data. Our out-of-sample results show that this computationally efficient technique surpasses conventional methods across risk metrics, risk-adjusted performance, and trading expenses in varied market conditions. Furthermore, in turbulent markets, it not only maintains but also enhances tracking performance.

\end{abstract}

\section{Introduction}

Portfolio optimization has been one of the most important topics in financial engineering and financial data science. In index tracking the idea is to replicate the performance of a stock market index, such as the S\&P 500 or the Dow Jones Industrial Average, by constructing a portfolio of securities that closely matches the index's return dynamics. This problem is of great interest to institutions that provide exchange-traded funds with passive strategies to obtain high diversification with low costs. Recently, data-driven approaches have been successfully developed for active and adaptive strategies on both index-tracking and mean-variance \cite{benidis2017sparse,baes2021low,li2022sparse} problems with sparse portfolios. In this paper, we present a computationally efficient, data-driven approach using Topological Data Analysis (TDA) that replicates the index with low tracking error, the small number of assets, and importantly, reduced risk. 

The problem of minimizing tracking error can be reformulated as a regression problem, where the optimal portfolio weights are represented by the estimated coefficients. To ensure the resulting portfolio's adaptability to market dynamics, it is crucial to incorporate index tracking as a form of parameter learning within an adaptive system. \cite{li2022sparse} In the recent literature, regularization techniques have been used to directly control the number of assets in sparse portfolios
with low tracking error, where sparseness is a desirable property from the point of view of transaction costs. The main concept involves imposing a constraint on the $\ell_q$-norm of portfolio weights to promote sparsity and stable allocations.{\footnote{Given a vector ${\bm w}=(w_1,w_2,\ldots,w_n)$, the $\ell_q$-norm is defined as $||{\bm w}||_{q} = (\sum_{i=1}^n|w_i|^q)^{\frac{1}{q}}$with $1\leq q<\infty$.}} For example, \cite{brodie2009sparse} suggests integrating an $\ell_1$-norm penalty on the vector of portfolio weights in optimization models to regularize the optimization problem and encourage sparse portfolios.{\footnote{The $\ell_1$-norm penalty, introduced to the statistical literature by  \cite{tibshirani1996regression}, is one of the most popular penalty. It is also known as the Least Absolute Shrinkage and Selection Operator (Lasso). It consists of an $\ell_1$-norm, whose unit ball resembles a cross-polytope with singularities at the coordinate axes. This characteristic makes it an appealing regularization method as it can address both variable selection (i.e., selecting a subset of assets for investment) and parameter estimation (i.e., determining the investment amount for the chosen assets) simultaneously, in a single step. However, in stark contrast, a recent study \cite{li2022sparse} employed the $\ell_0$-norm and decomposed the problem into two sequential subproblems: asset selection and capital allocation.}} Moreover, \cite{demiguel2009generalized} demonstrate that norm-constrained global minimum variance portfolio yields improved out-of-sample performance and \cite{giamouridis2010regular} and \cite{giuzio2018tracking} utilize $\ell_1$ penalty in the context of replicating hedge funds.

Recently, \cite{ho2015weighted} considered the mean-variance optimization with an advanced regularization technique, called weighted Elastic-Net, to overcome the limitations of mean-variance portfolio, such as the negative impact of estimation errors in mean and covariance on the out-of-sample performance. They obtained the parameters of the penalty terms from the estimation errors in the mean and the covariance matrices. In the context of this paper, the objective function for the portfolio tracking problem with the weighted Elastic-Net terms has a form of
\begin{equation}\label{eq:wElasticNet}
\sum_{t=1}^{T}(R_{{\bm w},t}-R_{0,t})^2 + \sum_{i=1}^n\alpha_i|w_i|+\sum_{i=1}^n\beta_{i}^2w_i^2,\\
\end{equation}
where $R_{0,t}$ is time $t$ return of an index to be tracked, $R_{{\bm w},t}$ is the return of the tracking portfolio, and $w_i$ is the weight of asset $i$. Moreover, the weight of a given asset $i$ is penalized by the regularization coefficients, $\alpha_i$ and $\beta_i$, $i = 1, 2, \dots, n$, that control the amount of regularization. However, in the index tracking problem, the covariance matrix is not solved in the first place, and for that reason, we cannot use the estimation approach of \cite{ho2015weighted} to estimate the parameter values of the penalty terms, $\{\alpha_i, \beta_i\}, i = 1, 2, \dots, n$. In this paper, the parameter values within the regularization terms in (\ref{eq:wElasticNet}) are learned through Topological Data Analysis (TDA){\footnote{Topological Data Analysis employs advanced mathematical concepts from algebraic topology to analyze complex, high-dimensional data and reveal hidden patterns \cite{carlsson2009topology, bubenik2015statistical, wasserman2018topological}. It is widely applied in biology \cite{lum2013extracting}, neuroscience \cite{geniesse2019generating}, and computer science \cite{pereira2015persistent}. In finance, TDA detects crash warnings \cite{gidea2017topological,gidea2018topological}, aids portfolio management and investment decision making \cite{baitinger2021better, goel2020topological}, and integrates with machine learning \cite{wu2022topological, ferri2018topology, moroni2021learning}, including a breakthrough neural network layer for persistence diagrams \cite{carriere2020perslay}. For an overview, refer to \cite{hensel2021survey} on topological machine learning methods.}}, employing a computationally efficient data-driven approach. To the best of our knowledge, this paper is the first to introduce an index tracking framework that incorporates a weighted Elastic-Net penalty, and it also represents a pioneering application of TDA to address the index tracking problem as a whole. This advancement is made possible by harnessing topological features, which facilitate the data-driven learning of coefficients. Although previous research has addressed specific cases of the regularization terms presented in equation (\ref{eq:wElasticNet}) for index tracking \cite{kremer2022sparse,shu2020high} and has demonstrated enhanced tracking performance through sparsity, either the $\ell_2$ term has been disregarded or it lacks asset-specific coefficients.

More specifically, in this paper, the regularization coefficients $\alpha_i$ and $\beta_i$ are obtained in a data-driven manner based on the TDA-based measures on the riskiness of asset $i$. While the primary objective of a sparse index tracking portfolio is to closely replicate the benchmark index's performance, it is crucial for the portfolio to be cost-effective and carry minimal risk. Therefore, the tracking portfolio should not only sufficiently track the index but also have smaller transaction costs and lower risk compared to the existing methods. Moreover, it is desirable to avoid expensive calibration schemes to obtain the parameter values. 

To use TDA to determine the weights in the regularization terms, we construct a point cloud of returns by transforming the return series from a given time window into a multi-dimensional form by arranging the observations into partially overlapping sub-intervals. If the point cloud of returns is highly concentrated (scattered), then the return observations between partially overlapped sub-intervals are close to (far from) each other, indicating return stability (variability). The level of concentration can be easily captured with TDA by calculating the $L_1$ norm of a so-called persistence diagram (PD).\footnote{It is important to avoid confusion between the $L_1$-norm of a persistence diagram and the $\ell_1$ regularization term in  (\ref{eq:wElasticNet}).} As a result, the lower (higher) $L_1$-norm of a PD, the higher the concentration (scattering) of the point cloud, and thus the more (less) stable the returns are over the sub-intervals from a topological point of view. By this way, the topological properties of the return data are learnt to capture the variability of returns from the topological point of view. One obvious alternative way to measure risk for the determination of the weights in the regularization terms is to estimate volatility from the asset's returns. This is used as a benchmark approach for the use of TDA. 

Our TDA approach has significant implications, the foremost being the ability to account for the ``topological risk'' associated with each asset. This is especially vital in financial modeling, where assigning less weight to higher norm values helps reduce the impact of potential outliers or extreme values, ultimately improving the model performance. Additionally, our approach incorporates topological features of underlying stocks into the modeling process, distinguishing it from traditional machine learning methods. Moreover, our approach is entirely data-driven. Last but not least, the regularization parameters can be learnt even from a smaller amount of data. This is different from regular methods which need a lot of data to estimate the regularization parameter effectively.

\section{Settings for Index Tracking with Sparse Portfolios}\label{section:settings}
Let $I$ be a stock-market index comprising $n$ assets $i\in\Ical=\{1,2,\ldots,n\}$, traded at discrete times $t=0,1,2,\ldots$, where $t$ represents the end of a trading day. The price of the index $I$ and any asset $i\in\Ical$ at time $t$ are denoted by $P_{0,t}$ and $P_{i,t}$, respectively. The return of asset $i$ over the period $[t-1,t]$ is denoted by $R_{i,t}$ and is given by
\begin{equation*}
R_{i,t}=\frac{P_{i,t}-P_{i,t-1}}{P_{i,t-1}},~~i \in \Ical\cup{0};~ t=1,2,\ldots.
\end{equation*}
with $i=0$ representing the return of the index. We define a portfolio as an $n\times 1$ vector, denoted by $\bm w=(w_{1},w_{2},\ldots,w_{n})$, where $w_{i} \in \mathbf{R}$ denotes the proportion of investment in the $i$th asset. At time $t$, the portfolio return $R_{{\bm w},t}$ is then given by the weighted sum of the individual asset's return, i.e.,
\begin{equation*}
R_{{\bm w},t}=\sum_{i=1}^{n}R_{i,t}w_i.  
\end{equation*}
The set of all admissible portfolios is denoted by $\Wcal$, 
\begin{equation*}
\Wcal=\left\{{\bm w}=(w_1,w_2,\ldots,w_n): \sum_{i\in \Ical} w_i=1,~~\forall i\in \Ical \right\}.
\end{equation*}

We assume that the investor holds the portfolio weights fixed until the end of a specific investment horizon. In the sparse index tracking problem, the portfolio weights $\bm w$ are selected so that 
\[
\sum_{t=1}^{T}(R_{{\bm w},t}-R_{0,t})^2 + \sum_{i=1}^n\alpha_i|w_i|+\sum_{i=1}^n\beta_{i}^2w_i^2,\\
\]
is minimized for given observations $\{R_{i,t}; i = 1, 2, \dots, n; t=1,2,\ldots,T\}$ and coefficients $\{\alpha_1, \alpha_2, \dots, \alpha_n\}$ and $\{\beta_1, \beta_2, \dots, \beta_n\}$. In this paper, we present an approach that extracts coefficients $\{\alpha_1, \alpha_2, \dots, \alpha_n\}$ and $\{\beta_1, \beta_2, \dots, \beta_n\}$ from risk measures with the use of TDA. As a benchmark, we use the following existing approaches:
\begin{itemize}
    \item We set $\alpha_i = \beta_i = 0$ for all $i = 1, 2, \dots, n$, i.e. no penalty functions are used (this is commonly referred to as Tracking Error, TE).
    \item We set $\beta_i = 0$ for all $i = 1, 2, \dots, n$ and 
    \begin{equation}\label{eq:slope} \sum_{i=1}^n\alpha_i|w|_{(i)}=\alpha_1|w|_{(1)}+\alpha_2|w|_{(2)}+\ldots+\alpha_n|w|_{(n)},
    \end{equation}     
    such that
    \[    \alpha_1\geq\alpha_2\geq\ldots\alpha_n~~\mbox{and}~~|w|_{(1)}\geq|w|_{(2)}\geq\ldots|w|_{(n)},
    \]
    where $[\alpha_1,\alpha_2,\ldots,\alpha_n]$ is a non-increasing sequence of tuning parameters and $|w|_{(i)}$ denotes the $i$th largest entry of the weight vector ${\bm w}$ in absolute value. This is commonly referred to as SLOPE. For more information, see \cite{kremer2022sparse}. 
    \item We set $\beta_i = \beta > 0$ for all $i = 1, 2, \dots, n$ and $\{\alpha_i; i = 1, 2, \dots, n\}$ are solved from the $\bm w$ that are optimized without a penalty. This is proposed in \cite{zou2009adaptive} and it is commonly referred to as adaptive Elastic-Net, EN.
\end{itemize}

The application of the above penalties in the context of index tracking has primarily been studied with the aim of obtaining sparse tracking portfolios and improved out-of-sample performance (\cite{wu2014nonnegative,wu2014nonnegative2,sant2020lasso,benidis2017sparse}). To overcome the limitations of Lasso ($\ell_1$), \cite{zou2006adaptive} introduced a weighted $\ell_1$ norm penalty, called adaptive Lasso penalty, given by $\sum_{i=1}^n\alpha_i|w_i|$, which was used by \cite{yen2014solving} and \cite{fastrich2015constructing} in the context of sparse mean-variance framework and by \cite{yang2016nonnegative} in the index tracking. Recently,  \cite{kremer2020sparse,kremer2022sparse} introduced the Sorted $\ell_1$ Penalized Estimator (SLOPE) to the portfolio selection problem. {\footnote{Researchers in the field have also explored non-convex penalties, including the Logarithmic Penalty (LOG), Smoothly Clipped Absolute Deviation (SCAD), and $\ell_q$-Norm (\cite{fastrich2015constructing} and \cite{giuzio2018tracking}). While these non-convex penalties are capable of generating clones with fewer active positions than Lasso, they are plagued by significant numerical problems and cannot guarantee convergence to the global minimum. Further,  \cite{kremer2020sparse,kremer2022sparse} demonstrated through empirical studies that the SLOPE penalty produces sparse tracking portfolios with low turnover and improved tracking statistics, often strategies that rely on non-convex LOG and SCAD LOG penalties.}}

\section{Proposed Method}

\subsection{Topological Data Analysis for Time-Series Data}
The idea is to first construct a simplicial complex from the point cloud $X=\{x_i\}^M_{i=1}$ in $\mathbb{R}^q$. A simplicial complex, $S$, is a finite set of simplices that satisfy two essential conditions: (i) any face of a simplex from $S$ is also in $S$, that is if $\omega\in S$ and $\sigma\subset \omega$ then $\sigma\in S$; and (ii) intersection of any two simplices in $S$ is either empty or shares faces.{\footnote{A $k$-simplex is a convex hull $[a_0, a_1, \ldots , a_k]=\{\sum_{i=0}^k\eta_ia_i: \sum_{i=0}^k\eta_i=1\}$ of $k+1$ geometrically independent points $\{a_0, a_1, \ldots , a_k\}$. Furthermore, the faces of a $k$-simplex $[a_0,\ldots,a_k]$ are the $(k - 1)$-simplices spanned by subsets of $\{ a_0,\ldots,a_k\}$. Note that 0-dimensional simplices represent data points (vertices), 1-dimensional simplices are connected pairs of vertices (edges), a 2-dimensional simplex is a filled triangle determined by its three vertices, and so on.}} There are many ways to construct a simplicial complex for a given point cloud and in this paper we adopt the Vietoris-Rips complex \cite{ghrist2008barcodes}. Specifically, the idea is to connect by edges the pairs of points (vertices) that are sufficiently close. Mathematically, we define 

\begin{defn}
Given a resolution parameter $\epsilon >0$, the Vietoris--Rips complex  of $X$ is defined to be  the simplicial complex $\Rcal_{\epsilon}(X)$ satisfying $[x_{i1},\ldots,x_{il}]\in \Rcal_{\epsilon}(X)$ if and only if $\text{Diam}(x_{i1},\ldots,x_{il})< \epsilon$. 
\end{defn}

Having obtained the simplicial complex, the next step is to extract the ``shape'' of the point cloud from it. TDA characterizes this shape through the presence of the topological features, for instance, the number of independent connected components or non-contractible loops. In fact, one can extract the higher dimensional topological features characterized by $k$-dimensional holes, where $k=0,1,2$ represents connected components, holes, and voids, respectively. These extracted topological features then serve as a proxy for the shape of the point cloud. 

The next question is how we define ``sufficiently close" (or choose $\epsilon$) when connecting points to obtain a simplicial complex. This is because the topological features in the complex depend on the choice of $\epsilon$. For instance, if $\epsilon$ is too small, we might see no edges, and hence no interesting topological characteristics. On the other hand, if $\epsilon$ is very large, then every point is connected to every other point leaving just one connected component, and consequently has no topological features of interest. TDA provides a principled solution for this by computing the shape of the point cloud over an entire range of resolution $\epsilon$ and studying the topological structure as a function of $\epsilon$ rather than making arbitrary choices that may skew our conclusions. More precisely, we obtain a sequence of Vietoris–Rips simplicial complexes, which we call Vietoris--Rips filtration and denote it by $\{\Rcal_{\epsilon_n}(X)\}_{n\in\mathbb{N}}$, for a non-decreasing sequence $\{\epsilon_n\} \in \mathbb{R}^+\cup\{0\}$ with $\epsilon_0= 0$. In this filtration, topological features would appear (birth) and disappear (death) in the corresponding simplicial complex, for instance, due to the addition of new edges or filling up of existing holes, etc. Each topological feature is given a `birth' and `death' value, and the difference between these values represents the feature's persistence. The information about the persistence of the topological features should provide a meaningful characterization of the overall shape of the data. For instance, features that persist over a wider range of scales are considered more robust, significant, and representative of the shape of the point cloud, while those that persist for a shorter range are deemed less significant or noisy and could be discarded if necessary. 

This process of studying the structure of complex data by analyzing its shape across multiple resolutions while simultaneously tracking the changes in its topological features is known as persistence homology. Figure \ref{rips} in the Appendix depicts this step-by-step procedure for obtaining the filtration. 

Taken's embedding (\cite{takens1981detecting}) is a standard procedure in topological analysis to embed a time-series into a high-dimensional space (point cloud) as TDA works on point-clouds in more than one dimension and 1-dimensional time-series do not naturally have point cloud representations 
(\cite{gidea2018topological,goel2020topological}). A time series $x=\{x_t,t=1,\ldots, T\}$ can be embedded to a multi-dimensional representation as 
    \begin{equation}\begin{split}
	{X} &= \left[
	\begin{array}{c}
	X_{1}\\
	X_{2}\\
	\vdots\\
	X_{T-(d-1)\tau}\\
	\end{array}
	\right]\\ 
    &=\left[
	\begin{array}{cccc}
	x_{1}& x_{1+\tau}\dots  &x_{1+(d-1)\tau}\\
	x_{2}& x_{2+\tau} \dots   &x_{2+(d-1)\tau}\\
	\vdots & \vdots & \vdots \\
	x_{T-(d-1)\tau} & x_{T-(d-2)\tau} \dots  &x_{T}\\
	\end{array}
	\right],\label{eq:point_cloud}
	\end{split}\end{equation}
where $\tau$ is the time delay, $d$ is the dimension of reconstructed space (embedding dimension), $T - (d - 1)\tau$ is the number of points (states) in the new space, and each point in the space is represented by a row of the matrix $X$. The delay embedding guarantees the preservation of topological features of a time series but not its geometrical properties. Even if it causes the loss of some geometric properties, it still has an important role. The point cloud is transformed into a filtration of simplicial complexes using the Vietoris-Rips filtration, as illustrated in Figure  \ref{rips} in the Appendix. This filtration involves drawing $\epsilon$ balls around each point in the point cloud and increasing $\epsilon$ gradually until all the balls merge. As this process occurs, various topological features, such as connected components, loops, and voids, emerge and disappear at different values of $\epsilon$.

One way to summarize the persistence of topological features in filtration is a persistence diagram (PD). A PD $\Dcal$ is  a  multi-set  of  points  in $\mathbb{W} \times \{ 0,1,\ldots,q - 1\}$,  where $\mathbb{W} :=\{(b,d)\in \mathbb{R}^2 : d\geq b\geq 0\}$ and each element $(b,d,f)$ represents a homological feature of dimension $f$ that appears at scale $b$ during a Vietoris--Rips filtration and disappears at scale $d$. Intuitively speaking, the feature $(b,d,f)$ is a $f$-dimensional hole lasting for duration $d - b$, called persistence. Namely, features with $f= 0$ correspond to connected components, $f= 1$ to loops, and $f= 2$ to voids. 

Though persistence diagrams contain potentially valuable information about the underlying data set, it is a multi-set which, when equipped with Wasserstein distance, forms an incomplete metric space and hence is not appropriate to apply tools from statistics and machine learning for processing topological features. Our aim in this paper is to study time series, and it is useful to embed the persistence diagrams into a Hilbert space to perform the requisite analysis.
One such embedding is the persistence landscape, introduced by \cite{bubenik2015statistical}. It comprises a sequence of continuous and piece-wise linear functions defined on a re-scaled birth-death coordinate, with the peaks of a persistence landscape representing the significant topological features. 
To obtain a persistence landscape, illustrated in Appendix in Figure \ref{fig:perlandscape}, the persistence diagram is rotated clockwise 45 degrees, and then treating the homology feature as a right angle vertex, isosceles right angle triangles are drawn from each feature. From the obtained collection of these triangles, individual functions are obtained. For instance, the third landscape function is the point-wise third maximum of all the triangles drawn. 
Mathematically, we have, for a fixed feature dimension $f$, we associate a piece-wise linear function $\Lambda _p(t) : \mathbb{R}\rightarrow [0,\infty)$ with each birth-death pair $p = (a, b) \in \Dcal$, where $\Dcal$ is the persistence diagram, as follows:
\begin{eqnarray}
\Lambda _p(t) = \left\{\begin{array}{ll} t-a &\quad t \in [a,\frac{a+b}{2}] \\[0.5em]
b-t &\quad t \in [\frac{a+b}{2},b]\\[0.5em]
0 &\quad \mbox{otherwise}.
\end{array}\right. \label{eq:perlandscape}
\end{eqnarray}
A persistence landscape of the birth-death pairs $p_i(a_i,b_i),\; i=1,\ldots,m,$ is the sequence of functions $\lambda: \mathbb{N}\times\mathbb{R}\rightarrow [0,\infty),$ as $\lambda(k,t) = \lambda_k(t)$ where $\lambda_k(t)$ denotes the $k$-th largest value of $ \{\Lambda _{p_i}(t),\; i=1,\ldots,m\}$. We set $\lambda_k(x) = 0$ if
the $k$-th largest value does not exist; so, $\lambda_k (t)= 0$ for $k > m.$ The persistence landscape representation enjoys several benefits over the persistence diagrams, \cite{ bubenik2015statistical, bubenik2018persistence, chazal2014stochastic, chazal2015subsampling, chazal2017introduction} such as they form 
a subset of the Banach space $L_p(\mathbb{N} \times \mathbb{R})$ consisting of sequences $\lambda= (\lambda_k)_{k \in \mathbb{N}}$. This set has an obvious vector space structure, and it becomes a Banach space when endowed with the norm which quantifies how ``spread out" or ``concentrated" the landscape functions are,

	\begin{align}
	||\lambda||_p=\left( \displaystyle \sum_{k=1}^\infty ||\lambda_k||^p_p \right)^{\frac{1}{p}},
	\end{align}
where $||\cdot||_p$ is the $L_p$-norm. The norm of a persistence landscape characterizes a point cloud by a single numerical value, and a higher norm value indicates the presence of more significant topological features in the underlying point cloud.

\subsection{TDA-based Regularization Coefficients}

To obtain the optimal regularization parameters for each asset, we adopt a strategy that involves utilizing the $L_p$ norm of persistence landscapes (PLs). The motivation to use $L_p$ norms of persistence landscapes corresponding to stock returns comes from empirical observations reported in numerous studies (see \cite{guo2020empirical,gidea2018topological,saengduean2020topological,goel2020topological}). The main concept is that when the point cloud of returns (as defined in Eq. \ref{eq:point_cloud}) is tightly clustered, it implies that observations on returns within partially overlapped sub-intervals are closely aligned. This suggests that returns exhibit consistent topological characteristics and lower variability. Conversely, if the point cloud is widely dispersed, the opposite holds true, indicating greater variability in returns. The concentration level of return observations can be effectively quantified using TDA through the computation of the $L_1$ norm of a persistence diagram, as denoted in Equation (\ref{eq:perlandscape}). Consequently, a lower (higher) $L_1$ norm of the persistence diagram corresponds to a higher concentration (scattering) of the point cloud. This, in turn, reflects the stability (instability) of returns over the sub-intervals, as observed from a topological perspective. Overall, the TDA norms can effectively track changes in the state of stock returns dynamics, thus motivating the use of $L_p$-norms as a measure of risk associated with the stock returns.

Building upon this, we incorporate persistence landscapes into the problem of partial index tracking, where the goal is to mimic the performance of the underlying index by investing in a limited number of stocks that constitute (not necessarily) the index. To be more precise, we propose to use the $L_1$ norm of the PL corresponding to the zero-dimensional features of asset $i$ as $\alpha_i$ and use the $L_1$ norm of the PL corresponding to the one-dimensional features of asset $i$ as $\beta_i$. Given our choice of $\alpha_i$'s and $\beta_i$'s, we construct the following TDA-based sparse index tracking models{\footnote{If the absolute value of weight is less than 10E-08, then we replace the weight with zero.}}: 
\begin{enumerate}
\item[1.] $\text{TDAEN12}$ portfolio: This portfolio corresponds to the weighted Elastic-Net model (\ref{eq:wElasticNet}) with $\alpha_i$ set equal to the $L_1$ norm of the Persistence Landscape corresponding to the zero-dimensional features of asset $i$, and $\beta_i$ as the $L_1$ norm of the Persistence Landscape corresponding to the one-dimensional features of asset $i$.

\item[2.] $\text{TDAEN11}$ portfolio: It is a simpler version of $\text{TDAEN12}$ portfolio with $\beta_i=\alpha_i$ for $i=1,\ldots,n$ and 
$\alpha_i$ set equal to the $L_1$ norm of the Persistence Landscape corresponding to the zero-dimensional features

\item[3.] $\text{TDA} \ell_1$ portfolio: This portfolio corresponds to the adaptive Lasso index tracking model where we set $\beta_i=0$ for $i=1,\ldots,n$, and the value of $\alpha_i$ is the $L_1$ norm of the Persistence Landscape corresponding to the zero-dimensional features of asset $i$. 
\end{enumerate}

More precisely, the regularization parameters $\alpha_i$ and $\beta_i$ for asset $i$ are obtained by Algorithm \ref{algo:main_algo}, which, in turn, uses Algorithms \ref{rips_algo} to compute Rips filtration, \ref{PD_algo} to compute persistence diagram $\Dcal_{X_i^{(j)}}$ for the filtration, and \ref{PL_algo} to compute persistence landscapes from the persistence diagrams. Moreover, Figure \ref{pd_pl2} in the Appendix gives a graphical representation of our approach. 

Our choice of adopting the mean landscape has several advantages. First, it offers a substantial computational speed-up as it reduces the computational complexity to compute the persistence landscape for a large time series by dividing it into computations over several sub-series. Second, it allows to account for heteroscedasticity and non-stationarity in the financial data. Third, it provides an online learning framework. More precisely, when the new data for the next month comes, we only need to find the persistence landscape corresponding to the data from the new sub-series and use the already calculated persistence landscapes from the previous training window (refer Section \ref{sec:empirical_results} for more details). In our analysis, we focus on zero- and one-dimensional topological features, i.e., the connected components and loops and we fix $p = 1$. 

\begin{algorithm}[!ht]
    \KwData{Time-series data $\{R_{i,t}\}_{t=1}^T$ of $M$ months, i.e., $T=21\times M$, for a given constituent $i \in \{1, 2, \dots, n\}$ of an index and a sequence of resolutions $\epsilon_0 < \epsilon_1 < \ldots < \epsilon_N$.}
    
    \KwResult{The regularization coefficients $\{\alpha_i, \beta_i\}$ for constituent $i \in \{1, 2, \dots, n\}$.}

+        Split time-series data into $\Tcal$ overlapping sub-series of length $\tilde{T}$ days with overlap of $h<\tilde{T}$ days such that $\Tcal=\frac{T-\tilde{T}}{h}+1$ (we use $M=6,~\tilde{T}=42$ and $h = \tilde{T}/2 = 21$)\;
        \For{$j = 0,1, 2, \dots, \Tcal-1$}{
            Extract the sub-series $\{R_{i,t}\}_{t=jh+1}^{jh+\tilde{T}}$
            Apply Takens' time-delay embedding with time delay $\tau$ and embedding dimension $d$ to obtain point cloud $X_{i}^{(j)}$ (we use $\tau=1$ and $d=3$)
{\scriptsize        
    \[
	X_{i}^{(j)}=
    \left[
	\begin{array}{cccc}
	R_{i,jh+1}& \dots  &R_{i,jh+1+(d-1)\tau}\\
	R_{i,jh+2}& \dots   &R_{i,jh+2+(d-1)\tau}\\
	\vdots & \vdots & \vdots \\
	R_{i,jh+\tilde{T}-(d-1)\tau} &\dots  &R_{i,jh+\tilde{T}}\\
	\end{array}
	\right]
	\]
 }
            
            Call algorithm \ref{rips_algo} to compute Rips filtration $\{\Rcal_{\epsilon_n}(X_{i}^{(j)})\}_{n\in\mathbb{N}}$ for the point cloud $X_{i}^{(j)}$ 
            Call algorithm \ref{PD_algo} to compute persistence diagram $\Dcal_{X_i^{(j)}}$ for the filtration $\{\Rcal_{\epsilon_n}(X_{i}^{(j)})\}_{n\in\mathbb{N}}$ 
            Call algorithm \ref{PL_algo} to compute persistence landscapes $\{\lambda_{i,0}^{(j)}(k)\}$ and $\{\lambda_{i,1}^{(j)}(k)\}$ from the birth-death pairs $\{(b^{(0)}_m, d^{(0)}_m)\}_{m=1}^{r_0}$ and $\{(b^{(1)}_m, d^{(1)}_m)\}_{m=1}^{r_1}$ corresponding to 0 and 1-dimensional components extracted from the persistence diagram $\Dcal_{X_i^{(j)}}$ 
      }
Compute the mean persistence landscapes $\{\overline{\lambda}_{i,0}(k)\}$ and $\{\overline{\lambda}_{i,1}(k)\}$ as follows 
\begin{equation*}
   \overline{\lambda}_{i,0}(k)= \frac{1}{\Tcal} \displaystyle \sum_{j=0}^{\Tcal-1} \lambda_{i,0}^{(j)}(k),~\overline{\lambda}_{i,1}(k)= \frac{1}{\Tcal} \displaystyle \sum_{j=0}^{\Tcal-1} \lambda_{i,1}^{(j)}(k)
\end{equation*} 

Compute the corresponding $p$-norm (we use $p=1$ and $k=1$)
\begin{equation*}\begin{split}
||\overline{\lambda}_{i,0}||_p &= \left( \displaystyle \sum_{k=1}^\infty ||\overline{\lambda}_{i,0}(k)||^p_p \right)^{\frac{1}{p}},\\
||\overline{\lambda}_{i,1}||_p &= \left( \displaystyle \sum_{k=1}^\infty ||\overline{\lambda}_{i,1}(k)||^p_p \right)^{\frac{1}{p}}
\end{split}\end{equation*}
 
Return $\alpha_i \leftarrow ||\overline{\lambda}_{i,0}||_p$ and $\beta_i \leftarrow ||\overline{\lambda}_{i,1}||_p$

\caption{Algorithm to determine regularization coefficients $\alpha_i$'s and $\beta_i$'s \label{algo:main_algo}}
\end{algorithm}


\begin{algorithm}[!ht]
    \KwData{A point cloud $X$ in $\mathbb{R}^d$, 
    \begin{align*}
    X^T = \left[X_1, X_2, \cdots, X_M \right],
	\end{align*}
where $X_i\in\mathbb{R}^d,~\forall i$ and a sequence of resolutions $ \epsilon_0 < \epsilon_1 < \ldots < \epsilon_N,$}
    
    \KwResult{Rips filtration $\{\Rcal_{\epsilon_n}(X)\}_{n=0}^N$.}

             \For{$n = 0,1, \dots, N$}{
             \For{$m = 1,2, \dots, M$}{
             Draw $B(X_m, \epsilon_n)$ \# a ball of radius $\epsilon_n$ around point $X_m$}
            $\Rcal_{\epsilon_n}(X)=\{[X_m,X_{m'}]|B(X_m, \epsilon_n) \cap B(X_m', \epsilon_n) \neq \emptyset\}$ \# $[X_m,X_{m'}]$ represents an edge between  $X_m$ and $X_{m'}$ 
             }
             \caption{Algorithm to obtain a Rips Filtration from a point cloud\label{rips_algo}}
\end{algorithm}

\begin{algorithm}[!ht]
    \KwData{A Rips filtration $\{\Rcal_{\epsilon_n}(X)\}_{n=0}^N$ and $s$ denoting the dimension of topological feature.}
    
    \KwResult{A persistence diagram $\Dcal_{X}$, a set of points representing birth and death times of $s$-dimensional topological features, i.e., $\Dcal_X=\{(b_1,d_1),(b_2,d_2),\ldots,(b_k,d_k)\}$}
                Initialize an empty persistence diagram $D_X$
             \For{$n = 0,1, \dots, N$}{ 
           
    Compute its $s$-dimensional homology $H_s(R_{\epsilon_n}(X))$ }
    $F:H_s(R_{\epsilon_1}(X))\mapsto H_s(R_{\epsilon_2}(X))$ for $\epsilon_1<\epsilon_2$ is the canonical map
    \begin{multline*}
    \begin{aligned}
    \Dcal_X = \{(\epsilon',\epsilon'')| &\text{ for any } s\text{-dim homology class } \alpha \text{ such that}\\
    i) &\text{~}\alpha \notin H_s(R_{\epsilon}(X)) \text{ for } \epsilon < \epsilon',\\ 
    ii) &\text{~for }\epsilon' \leq \epsilon < \epsilon'', \alpha   \in H_s(R_{\epsilon}(X)),\\
    iii) &\text{~}\alpha \notin H_s(R_{\epsilon''}(X)) \}
    \end{aligned}
    \end{multline*}
            
    
    \caption{Algorithm to obtain a persistence Diagram from a Rips Filtration \label{PD_algo}}
\end{algorithm}


\begin{algorithm}[!ht]
    \KwData{A list of birth-death pairs, $\{(b_i, d_i)\}_{i=1}^r$}
    
    \KwResult{$\{\lambda(k)\}$-the corresponding persistence landscape, a list of lists of critical points $(x,y)$.}
    Initialize $\Dcal=\{(b_i, d_i)\}_{i=1}^r$
  Sort $\Dcal$ first according to increasing $b$ and second according to decreasing $d$
            $k \gets 1$
            \While{$\Dcal \neq \emptyset$}{
            Initialize $\lambda(k)$
          Pop the first term $(b,d)$ from $\Dcal$; Let $p$ point to the next term 
         Add $(-\infty, 0), (b, 0), \left(\frac{b+d}{2}, \frac{d-b}{2}\right)$ to $\lambda(k)$ 
           \While{$Last(\lambda(k)) \neq (0,\infty)$}{
    \If{$d$ is maximal among remaining terms in $\Dcal$ starting at $p$}
  { Add $(d, 0), (\infty, 0)$ to $\lambda(k)$ }
    \Else
    { Let $(b', d')$ be the first of the terms starting at $p$ with $d' > d$ 
   Pop $(b', d')$ from $\Dcal$; Let $p$ point to the next term 
  \If{$b' > d$}
    { Add $(d, 0)$ to $\lambda(k)$ }
  \If{$b' \geq d$}
    {Add $(b', 0)$ to $\lambda(k)$} 
  \Else
    {Add $\left(\frac{b'+d}{2}, \frac{d-b'}{2}\right)$ to $\lambda(k)$ 
    Push $(b', d)$ into $\Dcal$ in order, starting at $p$; Let $p$ point to the next term }
  {Add $\left(\frac{b'+d'}{2}, \frac{d'-b'}{2}\right)$ to $\lambda(k)$ 
   $(b,d) \gets (b', d')$ }}
           }
           $k \gets k + 1$ }
  \textbf{Return} $\{\lambda(k)\}$
            
\caption{Algorithm to obtain a persistence landscape from a persistence diagram\label{PL_algo}}
\end{algorithm}

Following the above discussion, one could suggest simply using the asset volatility or standard deviation as the regularization parameter instead of using TDA norms. To settle this debate, we construct two additional portfolios namely $\text{Vol} \ell_1$ and $\text{VolEN}$ where we take the volatility of the assets as regularization parameters. In $\text{Vol}\ell_1$, we set $\alpha_i = \phi {\text{Vol}_i}$ and $\beta_i = 0$ for all $i = 1, 2, \dots, n$. In VolEN, we set $\alpha_i = {\phi}{\text{Vol}_i}$ and $\beta_i ={\psi} {\text{Vol}_i}$ for all $i = 1, 2, \dots, n$, where  
\begin{equation}\label{eq:volatility}
\text{Vol}_i = \sqrt{\frac{252 \times \sum_{t=1}^TR_{{\bm i},t}^2 }{T}} ~~i \in {1,2,\ldots,n},
\end{equation}
is the volatility of asset $i$. The parameters $\phi$ and $\psi$ are scaling parameters that are determined through a grid search process. The optimal values for these parameters are the ones that minimize the in-sample tracking error while also ensuring a comparable number of assets with non-zero weights as the TDA-based portfolio.

\section{Empirical results}\label{sec:empirical_results}

\subsection{Data}\label{data}

To investigate the performance of our proposed index tracking model, we focus on the daily closing prices of the S\&P500 and its constituents over a period spanning nearly 23 years from October 11, 1999, to March 8, 2023. The data is collected from Thomson Reuters Datastream. Since our goal is to obtain a low-risk tracking portfolio characterized by low tracking error and superior risk-return performance, we conduct a separate analysis of our model's performance during the recent COVID crisis period. To this end, we divide the data into two periods: October 11, 1999, to October 10, 2018, and January 1st, 2018, to March 8, 2023, encompassing 4957 and 1354 trading days respectively.

Since the index's composition is time-varying, we follow the standard approach from the literature to select the constituents \cite{goel2018index}. More precisely, we choose only those constituents for which the data is available for the whole period and drop the stocks with missing observations from our analysis. This results in 392 and 474 constituents in the first and second periods respectively.

\begin{table}[t!]
  \centering
  \caption{Descriptive statistics for the S\&P500 index for the two periods}
\scalebox{0.95}{\begin{tabular}{l|r|r}
    \hline
          & October 11, 1999 to& January 1st, 2018 to\\
          & October 10, 2018 & March 8, 2023 \\
          \hline
    Mean  & 6.730E-05 & 2.389E-04 \\
    Max& 4.759E-02 & 4.076E-02 \\
   Min& -4.113E-02 & -5.067E-02 \\
    Std Dev & 5.119E-03 & 7.901E-03 \\
    Volatility & 8.126E-02 & 1.254E-01 \\
    Skewness & -2.235E-01 & -3.675E-01 \\
    Kurtosis & 9.106 & 4.872 \\
    10th percentile & -5.428E-03 & -9.193E-03 \\
    50th percentile & 1.043E-04 & 2.983E-04 \\
   90th percentile & 5.135E-03 & 8.657E-03 \\
    \end{tabular}}%
  \label{tab:statistics}%
\end{table}%

\begin{figure*}[t!]
    \centering
    \includegraphics[width=18cm,height=21cm]{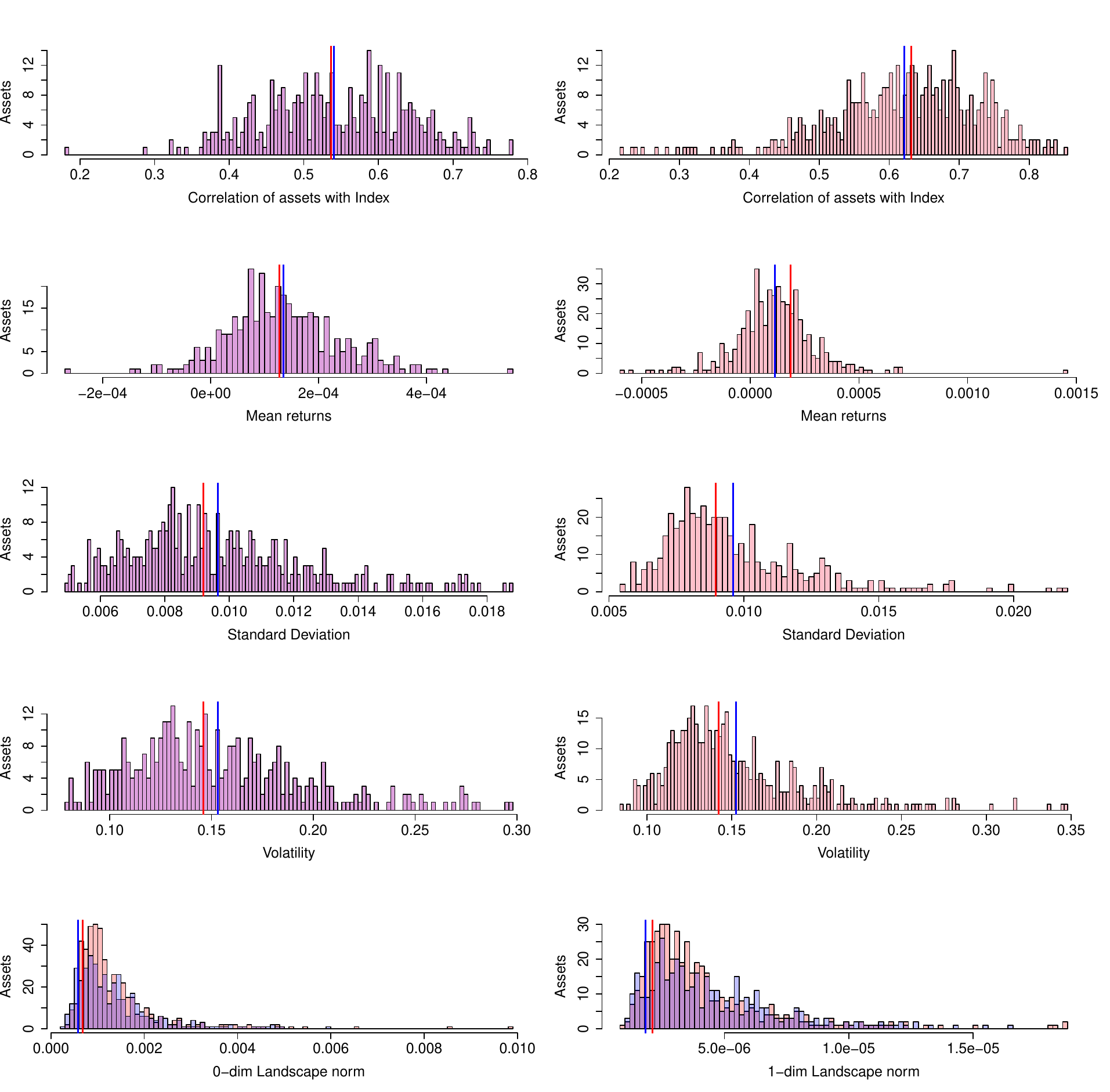}
    \caption{Correlation of assets with the index, mean returns, standard deviation, volatility, 0 and 1-dim Landscape norm for all the constituents for the period two periods. The blue (red) line in each of the histograms represents the corresponding median (mean) value of the statistic across all assets.}
    \label{fig:statistics}
\end{figure*}

Table \ref{tab:statistics} presents the descriptive statistics for the benchmark index S\&P500 during the two periods, confirming the typical stylized facts of financial time series, such as negative skewness (asymmetry) and high kurtosis (heavy tails). Further, as expected, it is evident that the standard deviation, (negative) skewness, and the minimum value are relatively larger in the COVID period. Due to the large number of constituents, presenting a complete table for descriptive statistics of each of the constituents is impractical. Therefore, we show the histograms in Figures \ref{fig:statistics} summarizing the correlation of constituents with the benchmark index, the mean return, the standard deviation, and the volatility, the 0-dimensional landscape norm, and the 1-dimensional landscape norm for constituents over the period of analysis. The histograms in the left Panel correspond to the first period while the right panel to the second period. The blue (red) line in each of the histograms represents the corresponding median (mean) value of the statistic.

\begin{figure*}[t!]
		\begin{center}
			\centering
			\includegraphics[scale=0.37]{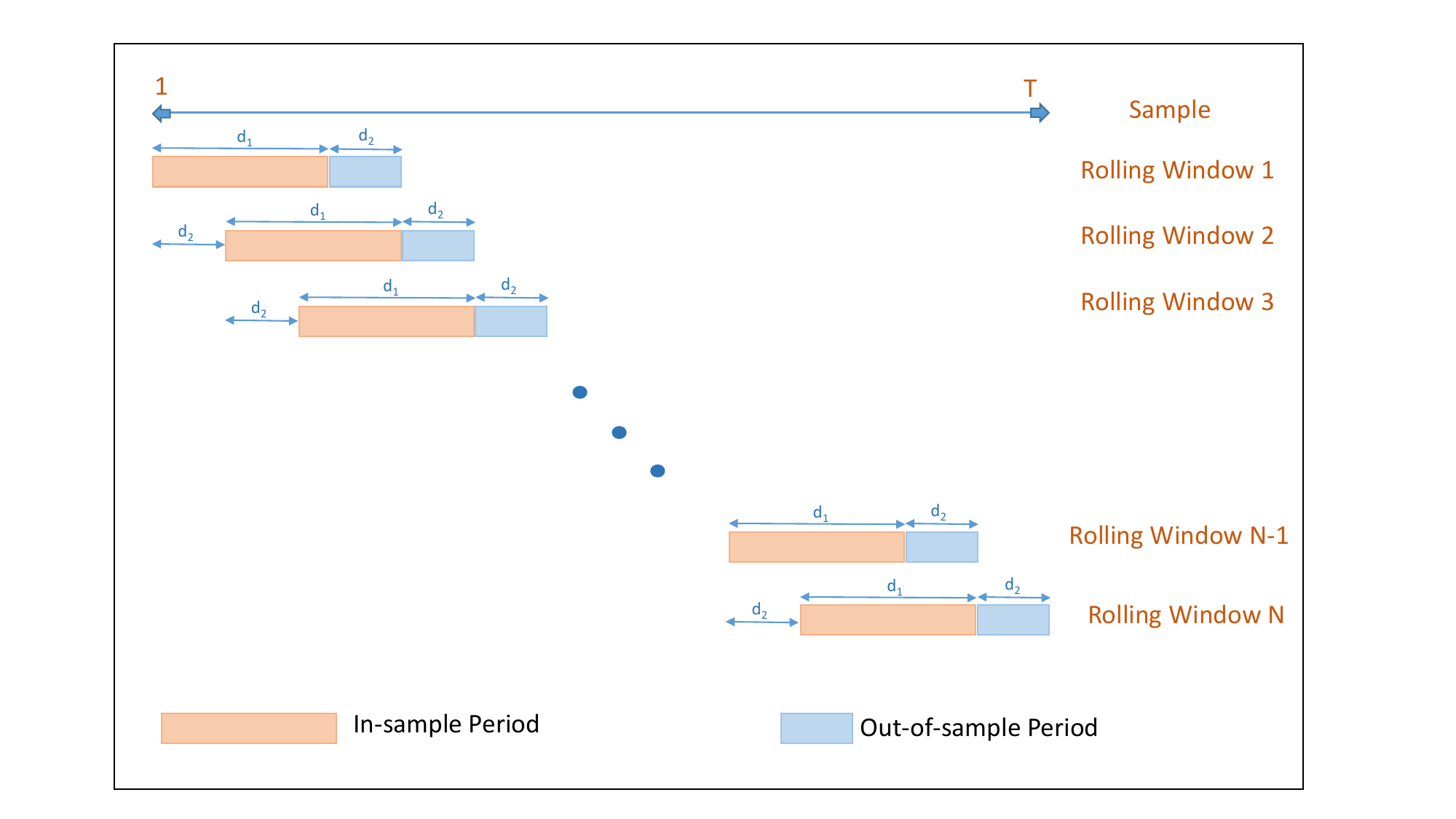}\\
			\caption{~An instance of a rolling window scheme is exemplified in this figure, where the orange and blue hues depict the in-sample and out-of-sample periods, respectively. Given a fixed in-sample period of length $d_1$ and an out-of-sample period of length $d_2$, the first in-sample period (rolling window 1) refers to the interval $1,\ldots,d_1$ with an associated out-of-sample period of $d_1 + 1,\ldots, d_1 + d_2$. The subsequent in-sample period (rolling window 2) is then shifted by $d_2$ to acquire a new in-sample window $1 + d_2,\ldots,d_1 + d_2$, with the corresponding out-of-sample period of $d_1 + d_2 + 1,\ldots,d_1 + d_2 + d_2$, and so forth. In this paper, the length of the in-sample periods $d_1 = 504$ trading days (25 months) and the length of the out-of-sample periods  $d_2 = 21$ trading days (one month). The first in-sample period is Oct 1999-Sept 2001 and the first out-of-sample is Oct 2001.} \label{fig:sliding}
		\end{center}
\end{figure*}

We utilize a rolling window approach to investigate the performance of our TDA-based sparse index tracking portfolio. The underlying idea of the rolling window approach is to gradually shift the in-sample (formation) and out-of-sample (tracking) period forward in time to include more recent observations and exclude the oldest observations such that a fixed size of the rolling window is maintained. We repeat this procedure until the end of the dataset. Specifically, at each rolling step, we first use the in-sample period to construct an optimal tracking portfolio by minimizing the in-sample tracking error subject to the respective penalty and budget constraints. We then track the performance of our portfolio over the new out-of-sample period, thus resulting in a sequence of performance measures. Figure \ref{fig:sliding} provides an illustration of the rolling window approach. Our analysis employs a rolling window of 525 trading days (25 months) with an in-sample period of 504 trading days (24 months) and an out-of-sample period of 21 trading days (one month). We shift the window by 21 trading days (one month) at each rolling step, leading to a total of 212 windows in the first period and 40 windows in the second period.


\subsection{Models and measures}

\subsubsection{Models and measures}

Next, we assess the performance of our proposed data-driven approach to determine regularization parameters. We create the following portfolios from the literature as benchmarks for comparative analysis (see Section \ref{section:settings}).
\begin{itemize}
\item[1.] $\text{TE}$ portfolio: This portfolio corresponds to a full replication index tracking portfolio, i.e., a tracking portfolio constructed using all the assets constituting the index, without penalty terms. 
\item[2.] $\text{SLOPE}$ portfolio: This corresponds to an index tracking portfolio with SLOPE penalty \cite{kremer2022sparse}. 
\item[3.] $\text{EN}$ portfolio: An index tracking portfolio with adaptive Elastic-Net penalty, originally proposed in \cite{shu2020high}.
\end{itemize}

To determine the regularization parameters for the SLOPE and EN models, we use the first 2 years of in-sample window data. We follow the approach outlined in \cite{kremer2022sparse} and \cite{shu2020high} respectively to learn the regularization parameter and set a range of parameters to ensure a comparable number of assets with non-zero weights as the TDA-based portfolio.

Though the goal of a sparse index tracking portfolio is to mimic the performance of the benchmark index closely, it is desirable that the portfolio is cost-effective, minimizes large losses, and is well-diversified in the sense that the exposure is not limited to a particular stock or a sector. Therefore, in our performance analysis, we include the metrics that measure i) tracking error (e.g., tracking error, correlation with the index), ii) risk measures, iii) risk-adjusted performance, and iv) turnover which reflects transaction costs
An optimal portfolio should have lower tracking error, risk-adjusted performance, small values of risk measures, and lower turnover indicating lower transaction costs.

Let $\Tcal$ denote the trading days in the out-of-sample period. We consider the following performance metrics in this paper.{\footnote{We refer the readers to \cite{bacon2023practical} for a comprehensive understanding of these measures.}} \\
\\
\textbf{i) Tracking performance measures}:
\begin{itemize}
    \item \textbf{Tracking error ($\text{TError}$):} It is the most commonly used metric by investors and portfolio managers to quantify the effectiveness of a portfolio's strategy in replicating the performance of a benchmark index. It is defined as 
    \begin{equation*}
    \textrm{TError} = \sqrt{\frac{\sum_{t\in\Tcal}(R_{{\bm w},t} - R_{0,t})^2}{|\Tcal|-1}}.    
    \end{equation*}
    A portfolio with a lower tracking error ($\text{TError}$) closely follows the benchmark, while a higher $\text{TError}$ indicates greater deviations from the benchmark's returns. Thus, a lower $\text{TError}$ is typically preferred.

    \item \textbf{Correlation:} It is Pearson's correlation coefficient between the returns of the benchmark index and the tracking portfolio. $correlation$ quantifies the degree of similarity between the two investment options.
\end{itemize}
\textbf{ii) Risk measures}:
\begin{itemize}
    \item \textbf{Annualized volatility:} See (\ref{eq:volatility}).

    \item \textbf{Downside Deviation ($\text{DD}$):} It provides a measure of the average deviation of those portfolio returns that fall below the benchmark return. It provides insight into the downside risk of a portfolio's returns and is given by 
    \begin{equation*}
    \textrm{DD}=\sqrt{\displaystyle\sum_{t\in\Tcal}\frac{(R_{0,t}-R_{{\bm w},t})^+)^2 }{|\Tcal|}}.
    \end{equation*}
    If the $\text{DD}$ value of a portfolio is lower, it indicates that the portfolio is generating the expected returns while limiting the potential downside risk. On the other hand, a higher $\text{DD}$ value suggests that the portfolio is consistently performing worse than the benchmark.

    \item \textbf{Value-at-Risk ($\text{VaR}_\alpha$):} It is a statistical measure used to estimate the maximum potential loss in a portfolio over a specific time horizon, at a given confidence level $\alpha\in(0,1)$. It is defined as
    \begin{equation}
	\textrm{VaR}_{\alpha}(-R_{{\bm w}}) = \min\{r \in \mathbb{R} \mid	F_w > \alpha\}\\
    \end{equation}
	\noindent where $F_w$ is the distribution for the portfolio loss $-R_{{\bm w}}$. In our analysis, we calculate the distribution of $R_{{\bm w}}$ using the time series of portfolio returns over the out-of-sample period. 	

    \item \textbf{Conditional Value-at-Risk ($\text{CVaR}_\alpha$):} 
    It is the expected value of portfolio losses beyond the $\text{VaR}_\alpha$  at a confidence level $\alpha\in(0,1)$. It is also known as Expected Shortfall (ES) or Tail Value-at-Risk (TVaR).

\end{itemize}
\textbf{iii) Risk-adjusted performance measures}:
\begin{itemize}
    \item \textbf{ Sharpe ratio (SHR)}: The Sharpe ratio is the average return earned
	in excess of the risk-free rate per unit of volatility as measured
	by the standard deviation, and is defined as:
	$$ \textrm{SHR} = \left\{\begin{array}{ll} \frac{\mu - R_f}{\sigma}, &\quad \mu > R_f \\[0.5em]
	0, &\quad  \mu \geq R_f,
	\end{array}\right.$$
	where $R_f$ denotes the risk-free rate, $\mu$ and $\sigma$ are the expected mean and standard deviation of the portfolio, respectively. We take $R_f=0$ for comparison.

    \item \textbf{ Information ratio (IR)}: It is the ratio of EMR to the tracking error, and is given by:
	$$ \textrm{IR} =  \frac{\textrm{EMR}}{\sigma_e},$$
    where excess mean return ($\text{EMR}$) is defined as the mean difference between the out-of-sample returns of the optimal tracking portfolio and the returns of the benchmark index, given by
    \[
    \textrm{EMR} = \frac{\sum_{t\in\Tcal}(R_{{\bm w},t}-R_{0,t})}{|\Tcal|},
    \]
    and $\sigma_e$ is the standard deviation of the excess return i.e. the difference between the portfolio returns and the benchmark returns.

    The higher values of the information ratio are preferable. Negative values of IR arise due to the negative value of EMR which indicates the under-performance of the portfolio with respect to the benchmark.
\end{itemize}
\textbf{iv) Turnover measure}:
\begin{itemize}
    \item \textbf{Turnover Ratio ($\text{TR}$):} It is the average absolute value of trades among $n$ stocks, defined by
	$$\text{TR}=\frac{1}{M} \sum _{r=1}^{M}\sum_{i=1}^{n}|w^r_{i}-w^{r-1}_{i}|,$$ where $M$ is the total number of re-balancing windows, $w^r_{i}$ and $w^{r-1}_i$ are weights of $i$-th asset in $r$-th and $(r-1)$-th windows, respectively. The smaller values of TR are desirable on account of lower transaction costs.
\end{itemize}

\subsection{Results}\label{subsection:results}

Panel I in Table \ref{tab:performance_analysis} shows the out-of-sample performance{\footnote{As we have a considerable number of windows within each data period to provide comprehensive details, we present the results based on the series created by concatenating the out-of-sample returns from each optimal in-sample portfolio.}} of tracking portfolios for data from October 11, 1999, to October 10, 2018 and Panel II from the Covid-19 January 1st, 2018 to March 8, 2023. Both Panels report results for models with two penalty terms, $\ell_1$ and $\ell_2$ (see Sub-Panels Ia and IIa), and models with one penalty term,  $\ell_1$ (see Sub-Panels Ib and IIb). We also report the performance of the benchmark full replication portfolio without any constraints on the norm of portfolio weights, denoted by $\text{TE}$.

\begin{table*}[t!]
\centering
\caption{The out-of-sample performance for different sparse index tracking portfolios. Panel I reports results for data from October 11, 1999, to October 10, 2018, and Panel II from the Covid-19 January 1st, 2018 to March 8, 2023. The first column shows the results for the non-penalized $\text{TE}$ portfolio. Within both Panels, we report results for models that use both $\ell_1$ and $\ell_2$ penalty terms, i.e. Elastic-Net type penalty models, namely $\text{TDAEN11}$, $\text{TDAEN12}$, $\text{VolEN}$, and $\text{EN}$ (see Panels Ia and IIa). Moreover, results for models based on $\ell_1$ penalty term only, namely $\text{TDA} \ell_1$, $\text{Vol} \ell_1$, and $\text{SLOPE}$ with Lasso type penalty are reported (see Panels Ib and IIb). The best values amongst the Lasso type and Elastic-Net type penalties are highlighted in \textbf{bold} separately, and the best among all the values have been \textit{italicized}. Rather than reporting the results of each out-of-sample period (see Figure \ref{fig:sliding}), we concatenate the out-of-sample returns of each period.}
\scalebox{.78}{%
\begin{tabular}{l|r|rrrr|rrr}
\hline
\multicolumn{9}{c}{Panel I:  Period 10/1999--10/2018}\\
\hline
&   & \multicolumn{4}{c|}{Panel Ia: $\ell_1$ and $\ell_2$ penalty} & \multicolumn{3}{c}{Panel Ib: $\ell_1$ penalty, 10/1999--10/2018} 
          \\
\hline
& \multicolumn{1}{l|}{TE} & \multicolumn{1}{l}{$\text{TDAEN11}$} & \multicolumn{1}{l}{$\text{TDAEN12}$} & \multicolumn{1}{l}{$\text{VolEN}$} & \multicolumn{1}{l|}{$\text{EN}$} & \multicolumn{1}{l}{$\text{TDA} \ell_1$} & \multicolumn{1}{l}{$\text{Vol} \ell_1$} & \multicolumn{1}{l}{$\text{SLOPE}$} \\
\hline
\textbf{Tracking performance:} & & & & & & & & \\
$\text{\quad TError}$ & 7.542E-04 & 6.368E-04 & 6.193E-04 & 6.197E-04 & \textit{\textbf{5.391E-04}} & \textbf{6.502E-04} & 9.484E-04 & 1.003E-03 \\
$\text{\quad Correlation}$ & 98.9\% & 99.2\% & 99.3\% & 99.1\% & \textbf{99.4\%} & \textbf{99.2\%} & 99.0\% & 98.1\% \\
\textbf{Risk:} & & & & & & & & \\
$\text{\quad Volatility}$ & 8.207E-02 & 7.799E-02 & \textit{\textbf{7.792E-02}} & 7.864E-02 & 8.110E-02 & \textbf{7.810E-02} & 7.866E-02 & 8.156E-02\\
$\text{\quad DD}$ & 5.553E-04 & 4.334E-04 & \textit{\textbf{4.212E-04}} & 4.380E-04 & 4.845E-04 & \textbf{4.433E-04} & 5.586E-04 & 7.549E-04 \\
$\text{\quad VaR}_\alpha$ & 7.849E-03 & 7.364E-03 & \textbf{\textit{7.345E-03}} & 7.389E-03 & 7.675E-03 & \textbf{7.369E-03} & 7.910E-03 & 7.670E-03 \\
$\text{\quad CVaR}_\alpha$ & 1.293E-02 & 1.207E-02 & \textbf{\textit{1.206E-02}} & 1.219E-02 & 1.270E-02 & \textbf{1.209E-02} & 1.919E-02 & 1.288E-02 \\
\textbf{Risk-adjusted performance:} & & & & & & & & \\
$\text{\quad SHR}$ & 1.147E-02 & 2.031E-02 & \textbf{\textit{2.033E-02}} & 1.733E-02 & 1.480E-02 & \textbf{2.016E-02} & 1.738E-02 & 7.700E-03 \\
$\text{\quad IR}$ & -4.779E-02 & 8.936E-03 & \textbf{\textit{9.261E-03}} & -1.370E-02 & -3.599E-02 & \textbf{7.724E-03} & -1.263E-02 & -5.553E-02 \\
\textbf{Turnover:} & & & & & & & & \\
$\text{\quad TR}$ & 1.135 & 0.224 & \textbf{\textit{0.206}} & 0.429 & 0.514 & \textbf{0.238} & 0.560 & 0.239 \\
\textbf{\# Assets} & 392 & 113 & 124 & 127 & 128 & 106 & 111 & 103 \\
\hline

\hline
\multicolumn{9}{c}{Panel II:  Period 1/2018--3/2023}\\
\hline
& &\multicolumn{4}{c|}{Panel IIa: $\ell_1$ and $\ell_2$ penalty} & \multicolumn{3}{c}{Panel IIb: $\ell_1$ penalty} \\
\hline
& \multicolumn{1}{l|}{TE} & \multicolumn{1}{l}{$\text{TDAEN11}$} & \multicolumn{1}{l}{$\text{TDAEN12}$} & \multicolumn{1}{l}{$\text{VolEN}$} & \multicolumn{1}{l|}{$\text{EN}$} & \multicolumn{1}{l}{$\text{TDA} \ell_1$} & \multicolumn{1}{l}{$\text{Vol} \ell_1$} & \multicolumn{1}{l}{$\text{SLOPE}$} \\
\hline
\textbf{Tracking performance:} & & & & & & & & \\
$\text{\quad TError}$  & 3.376E-03 & 1.014E-03 & \textbf{\textit{9.802E-04}} & 1.138E-03 & 1.177E-03 & \textbf{1.031E-03} & 1.063E-03 & 1.127E-03 \\
$\text{ \quad Correlation}$ & 87.9\% & 98.8\% & \textbf{\textit{98.9\%}} & 98.8\% & 98.3\% & \textbf{98.7\%} & 98.5\% & 98.4\% \\
\textbf{Risk:} & & & & & & & & \\
$\text{\quad Volatility}$ & 1.124E-02 & 9.091E-03 & \textbf{\textit{9.079E-03}} & 1.232E-02 & 1.267E-02 & \textbf{9.144E-03} & 1.220E-02 & 1.227E-02\\
$\text{ \quad DD}$ & 2.513E-03 & \textbf{\textit{6.436E-04}} & 6.133E-04 & 7.099E-04 & 8.654E-04 & \textbf{6.652E-04} & 8.115E-04 & 8.836E-04 \\
$\text{ \quad VaR}_\alpha$ & 9.337E-03 & \textbf{\textit{8.431E-03}} & 8.469E-03 & 8.852E-03 & 8.830E-03 & \textbf{8.561E-03} & 8.749E-03 & 9.565E-03 \\
$\text{ \quad CVaR}_\alpha$ & 1.733E-02 & \textbf{\textit{1.425E-02}} & 1.423E-02 & 1.552E-02 & 1.581E-02 & \textbf{{1.430E-02}} & 1.558E-02 & 1.564E-02 \\
\textbf{Risk-adjusted performance:} & & & & & & & & \\
$\text{ \quad SHR}$ &  2.350E-02 & 3.615E-02 & \textbf{\textit{3.624E-02}} & 2.257E-02 & 2.909E-02 & \textbf{3.573E-02} & 2.229E-02 & 1.288E-02 \\
$\text{ \quad IR}$ & -6.653E-03 & 2.651E-02 & \textbf{\textit{2.775E-02}} & -4.631E-02 & 4.470E-04 & \textbf{2.435E-02} & -4.216E-02 & -9.133E-02 \\
\textbf{Turnover:} & & & & & & & & \\
$\text{ \quad TR}$ & 4.589 & 0.389 & \textbf{\textit{0.356}} & 0.698 & 0.737 & {\textbf{0.423}} & 0.746 & {0.458} \\
\textbf{\# Assets} & 474 & 69 & 83.5 & 80.5 & 63 & 59 & 59 & 55 \\
\hline

\end{tabular}
}
\label{tab:performance_analysis}
\end{table*}

Our analysis reveals several key findings. Let us analyze the tracking performance first. In Panel I, which shows results for a longer period spanning from 10/1999 to 10/2018, the tracking performance is the best with EN portfolio and the worst with the SLOPE in terms of both tracking error and correlation with the index. Nonetheless, in terms of correlation, the difference in the performance is comparable and marginal (99.4\% vs 99.3\%) to a TDA-based method (TDAEN12). However, when it comes to the tracking performance in the more volatile period from 1/2018--3/2023 reported in Panel II, which includes the Covid-19 pandemic, TDA-based methods proposed in this paper clearly outperform the existing EN and SLOPE methods with a clear difference (VolEN and SLOPE have 20\% and 15\% higher tracking error and 0.6 and 0.5 percentage point higher correlation). This is not entirely surprising, because intuitively, the more important it is to take risk measures into account, the more markets fluctuate.

Our analysis has unveiled several significant findings, starting with an examination of tracking performance. In Panel I, encompassing data from October 1999 to October 2018, the EN portfolio demonstrates the strongest tracking performance, while the SLOPE portfolio shows the poorest, as evidenced by both tracking error and correlation with the index. Nevertheless, in terms of correlation, the difference of EN to the TDA-based methods in performance is relatively marginal, with TDA-based method TDAEN12 closely trailing at 99.3\% compared to the 99.4\% achieved by EN. However, when evaluating tracking performance during the more turbulent period spanning from January 2018 to March 2023, as presented in Panel II, which includes the COVID-19 pandemic, the TDA-based methods introduced in this paper significantly outperform the existing EN and SLOPE methods. This is demonstrated with significant gaps, with VolEN and SLOPE exhibiting 20\% and 15\% higher tracking errors and 0.6 and 0.5 percentage points lower correlation, respectively. This outcome aligns with intuition, as during periods of greater market volatility, the importance of incorporating risk measures becomes more pronounced due to heightened market fluctuations.

Secondly, our aim is to attain not only good tracking performance but also a low-risk level. In terms of all the four risk measures (Volatility, Downside Deviation, Value-at-Risk, and Conditional Value at Risk), Panel I provides clear evidence of the superiority of TDA-based measures across all four metrics: $\text{TDAEN12}$ consistently yields the lowest risk measures when compared to other models using both $\ell_1$ and $\ell_2$ penalties, whereas $\text{TDA} \ell_1$ stands out as the top performer among models employing $\ell_1$ penalty. In fact, the contrast compared to traditional methods (EN and SLOPE) is quite substantial. Panel II (Covid-19 period) further confirms that TDA-based approaches $\text{TDAEN11}$ and $\text{TDAEN12}$ outperform the traditional methods based on $\ell_1$ and $\ell_2$ penalties. Moreover, $\text{TDA} \ell_1$ again remains the superior method among models based on $\ell_1$ penalty. Overall, this shows that introducing a risk-component in the estimation procedure of asset weights clearly reduces the risk of the tracking portfolio, and the TDA-approaches clearly outperform the use of volatility in this regard. 

Thirdly, the TDA-based approaches excel in terms of risk-adjusted performance, too, as evidenced by the Sharpe and Information Ratios. This is consistent across both the 1999-2018 and 2018-2023 periods, suggesting that the use of TDA can deliver superior performance under varying market conditions. It is crucial to note that while the risk-adjusted performance of the tracking portfolio is noteworthy, our primary goal is to minimize the tracking error while managing the level of risk. This performance is more of a beneficial byproduct, further underscoring the efficacy of the methods proposed in this paper.

Fourthly, the results show that the turnover (defined as the average absolute value of trades) with $\text{TDAEN12}$ is less than half of that with $\text{EN}$, making it the method with the lowest turnover among all examined. Additionally, $\text{TDA} \ell_1$ has a lower turnover than SLOPE, whereas $\text{VOL} \ell_1$ performs the poorest in this aspect. On the whole, employing TDA leads to the most reduced transaction costs in portfolio tracking. It is worth noting that the number of assets does not vary very much between TDA-based methods and traditional approaches.

\begin{figure*}[ht!]
		\begin{center}
			\subfigure[Period 10/1999--10/2018]{%
				\includegraphics[height=9.5cm,width=14cm]{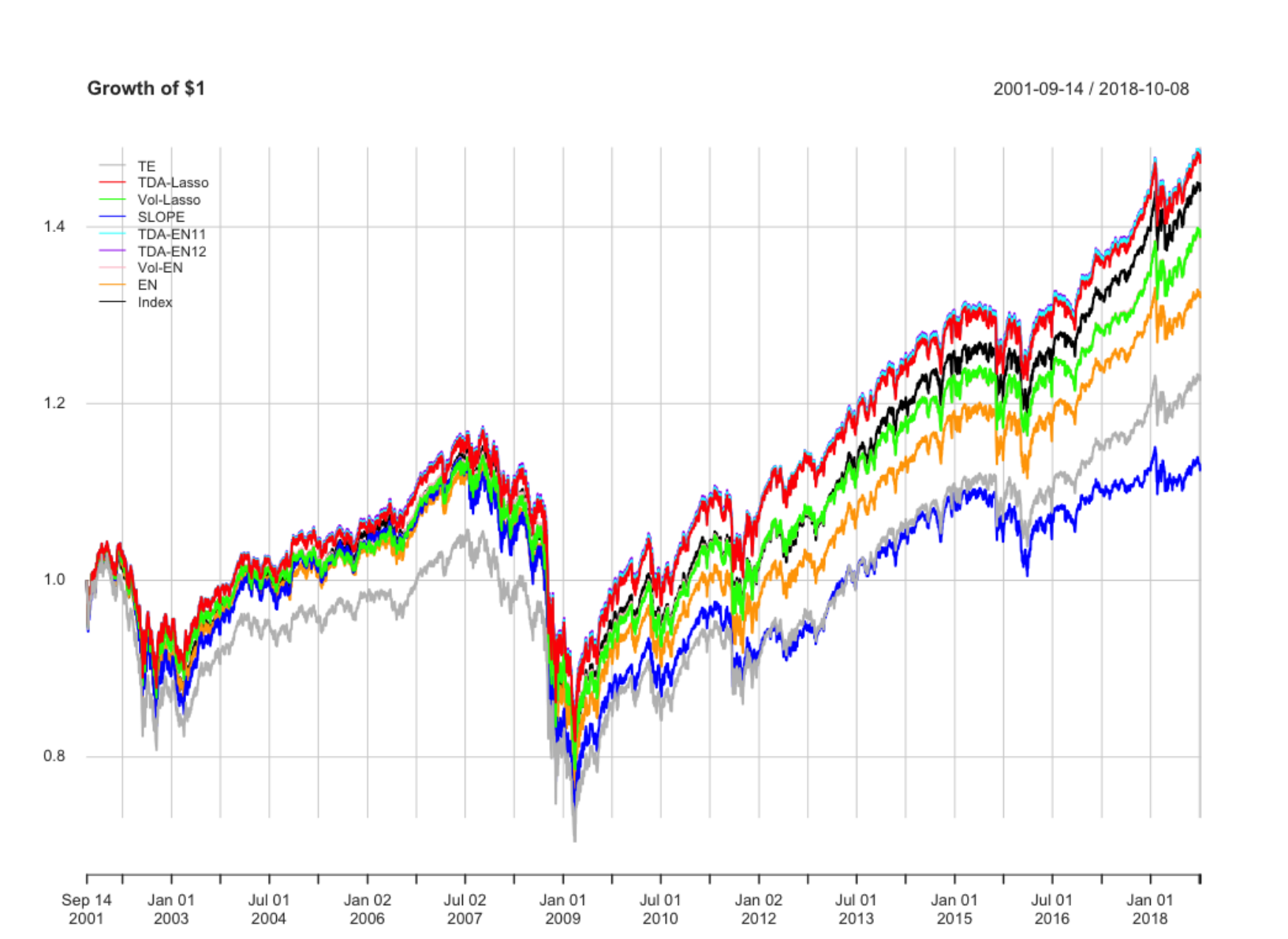}
				\label{fig:performance_analysis_wealth}}
			\quad
			\subfigure[ Period 1/2018-3/2023 (COVID-19)]{%
				\includegraphics[height=9.5cm,width=14cm]{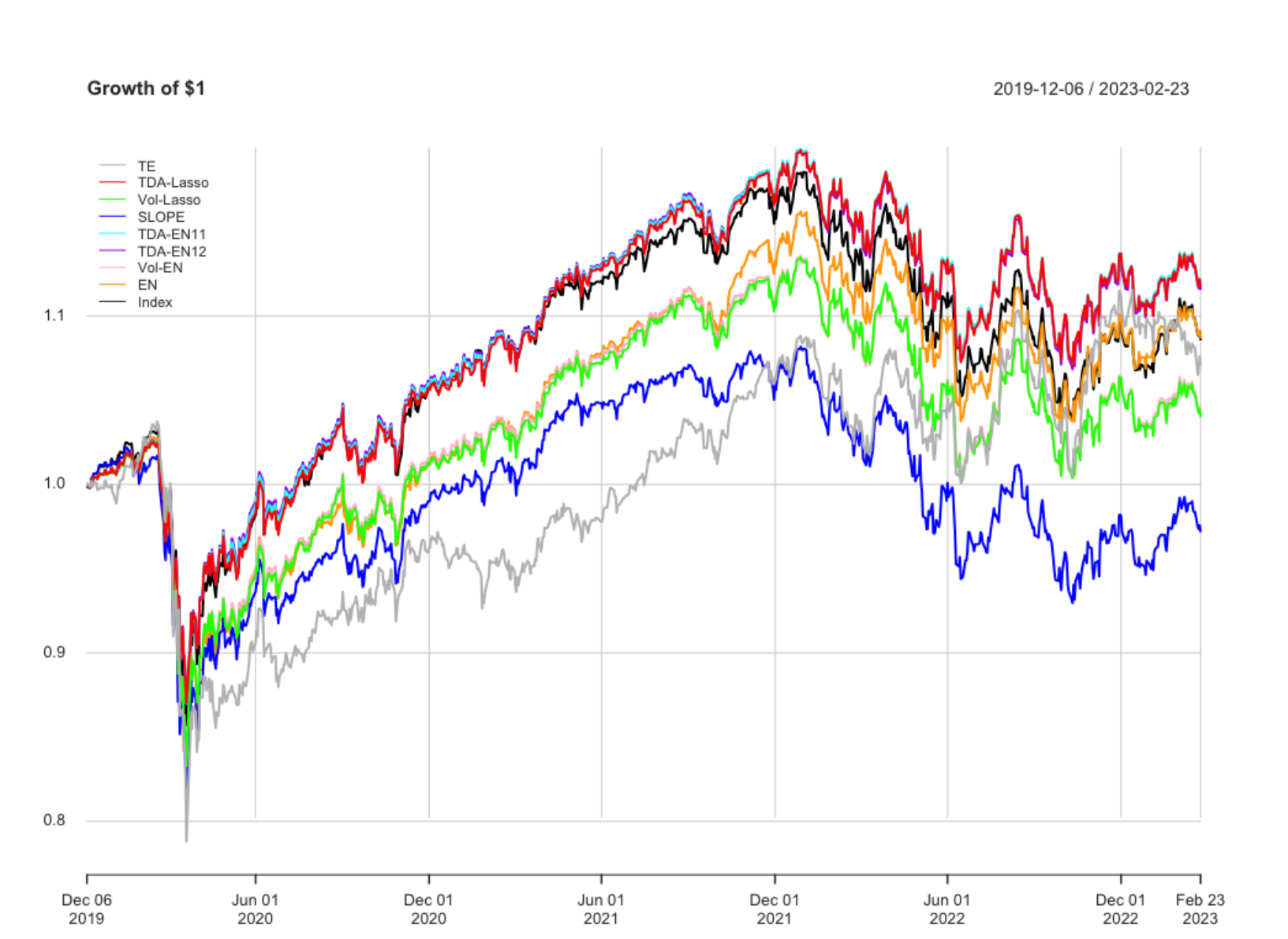}
				\label{fig:performance_analysis_covid_wealth}}
			\end{center}
		\caption{The wealth of the portfolios starting from \$1. Plot (a) shows the cumulative returns for period 10/1999--10/2018 (with 10/1999--9/2001 being the first in-sample period) and plot (b) for period 1/2018--3/2023 (with 1/2018--12/2019 being the first in-sample period).}\label{figure:performance_analysis}
\end{figure*}

Finally, we visualize the dynamics of the index and the tracking portfolios obtained with different methods. Figure \ref{figure:performance_analysis} shows the growth of \$1 investment (a) from September 2001 to January 2018 and (b) from December 2019 to February 2023. The corresponding two-year in-sample periods are from October 1999 to September 2001 and January 2018 to December 2019. As depicted in Figure \ref{figure:performance_analysis}, it is evident that the portfolios constructed using TDA-based regularization parameters not only efficiently track the benchmark index but also safeguard them against significant losses. These results underline the out-of-sample efficacy of TDA-based models in managing risk and delivering returns. 

The in-sample performance of the portfolios in both the periods considered is presented in the Appendix. These in-sample results, combined with out-of-sample results reported in Table \ref{tab:performance_analysis} show that TDA-based penality methods effectively manages overfitting in portfolio tracking problems. In summary, TDA-based methods, particularly $\text{TDAEN12}$, produce a low tracking error (with the lowest error observed in the Covid-19 data). Simultaneously, they diminish risk, enhance risk-adjusted performance, and cut down transaction costs for the tracking portfolio, which are highly desirable qualities.

\section{Conclusions}

In this paper, we present a novel data-driven approach to the index tracking problem with sparse portfolios, leveraging Topological Data Analysis (TDA). Our method eliminates the need for costly estimation procedures and inherently manages the risk of the tracking portfolio. Unlike traditional sparse portfolio techniques for index tracking, our approach accommodates both $\ell_1$ and $\ell_2$ penalty terms in the objective function (see Eq. \ref{eq:wElasticNet}). Additionally, our approach employs TDA-based risk measures to explicitly control the riskiness of the tracking portfolio. 

Our empirical findings demonstrate that the tracking error of our approach is not merely comparable, but often excels in turbulent market conditions relative to the prevailing methods. Furthermore, our findings show the superiority of our TDA-based approach across multiple metrics: the risk associated with the tracking portfolio is reduced, risk-adjusted performance is enhanced, and turnover (indicative of transaction costs) is clearly lower. Put simply, this study affirms that incorporating a TDA-based risk component during the asset weight estimation process substantially diminishes the risk of the tracking portfolio. 

\bibliography{Main_IEEE.bib}
\bibliographystyle{ieeetr}


\section*{Appendix}

\subsection{Graphical illustrations}\label{sec:graphical_illustrations}

\begin{figure}[!ht]
		\begin{center}
  			\subfigure[Time series]{%
				\includegraphics[height=4cm,width=6cm]{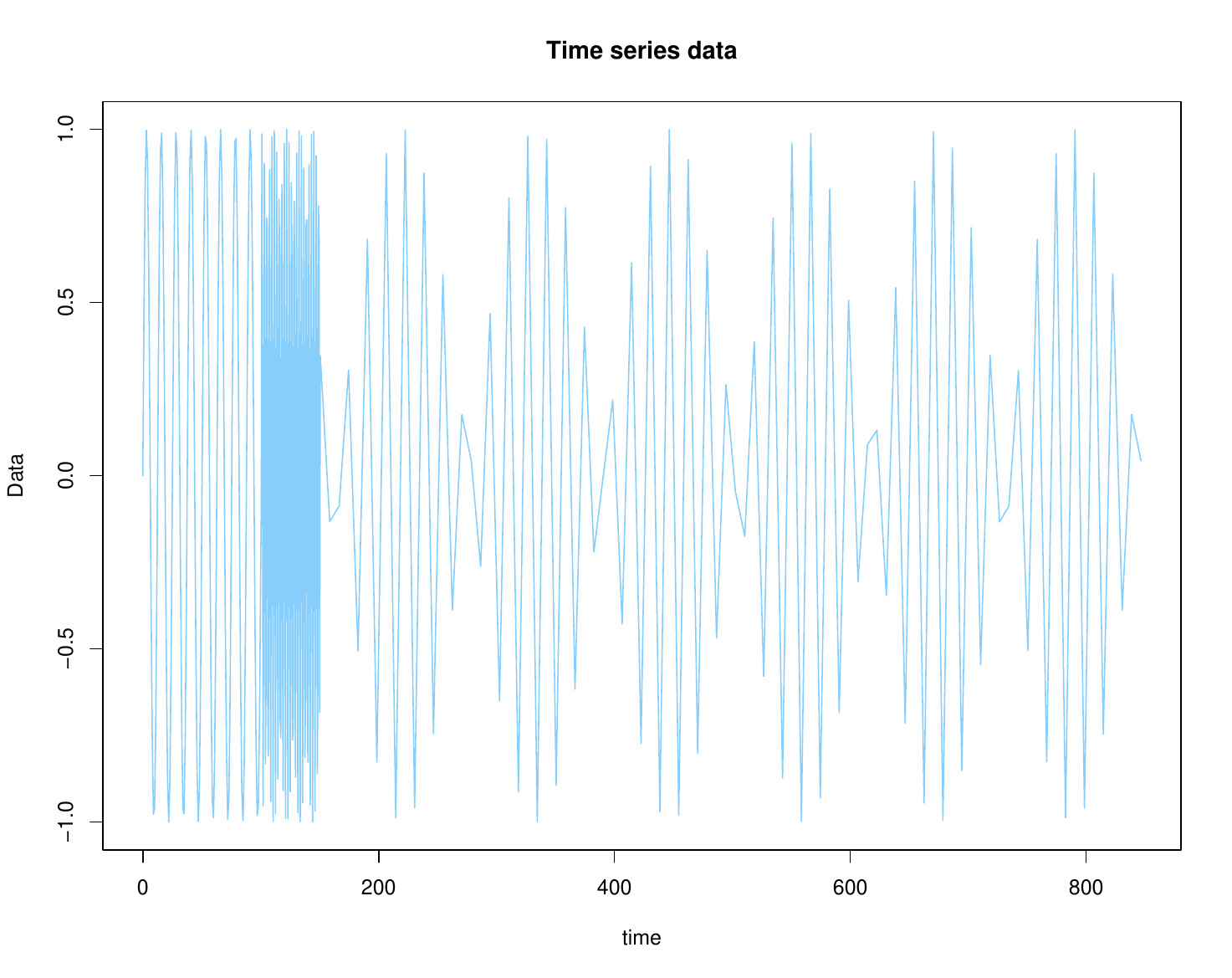}
				\label{fig:cloud_timeseries}}
    \quad
			\subfigure[Point cloud]{%
				\includegraphics[height=4cm,width=6cm]{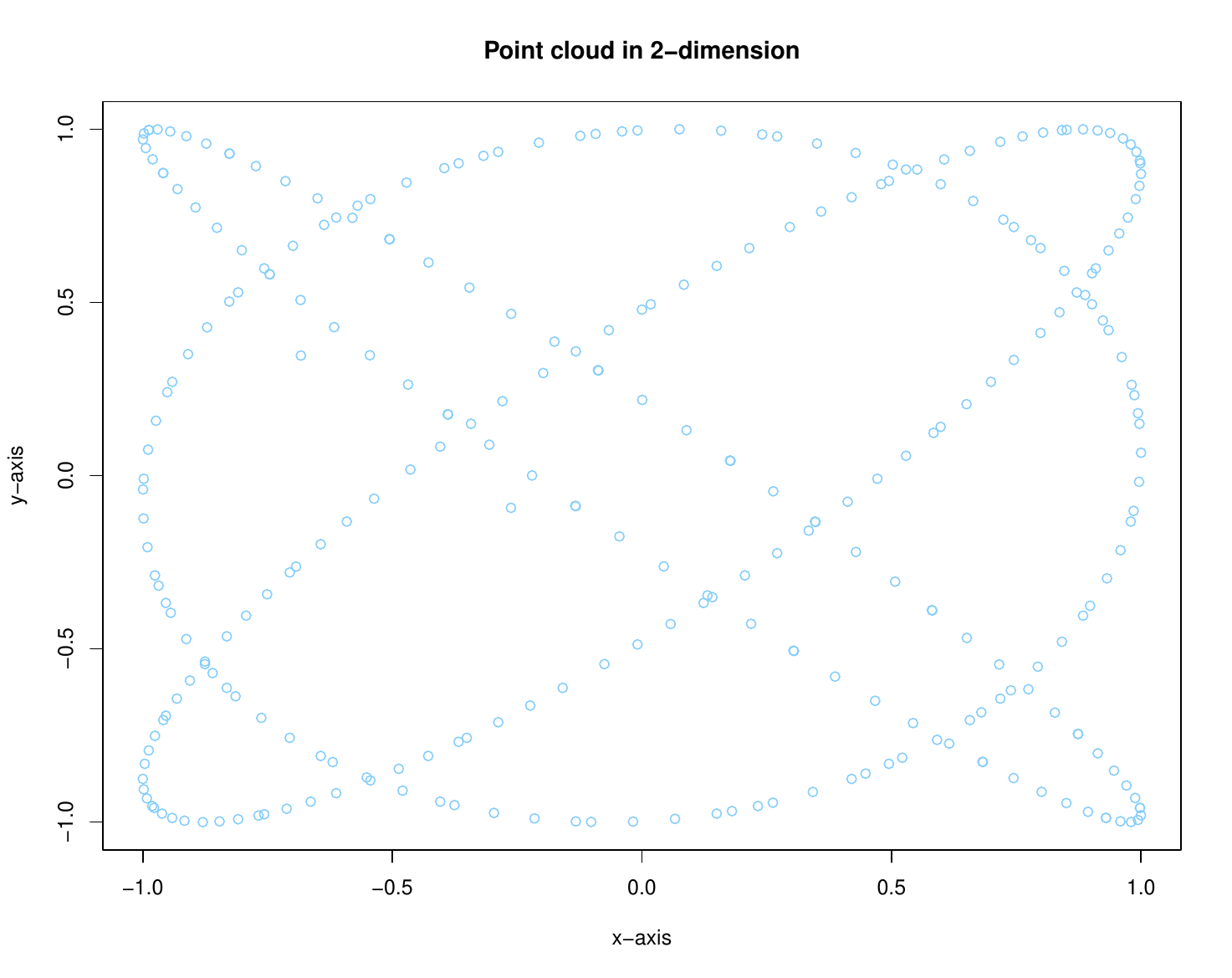}
				\label{fig:cloud_pointcloud}}
		\end{center}
		\caption{ A diagram depicting the Rips filtration process. Plot (a) displays the time series, while plot (b) shows the corresponding 2-dimensional point cloud, obtained using Taken's embedding with $d=2$ and $\tau=1$. 
        T} \label{fig:cloud}
	\end{figure}

\begin{figure}[!ht]
		\begin{center}
			\subfigure[$\epsilon_1=0.07$, $\text{loops}=13$]{%
			\includegraphics[height=4.5cm,width=0.45\textwidth]{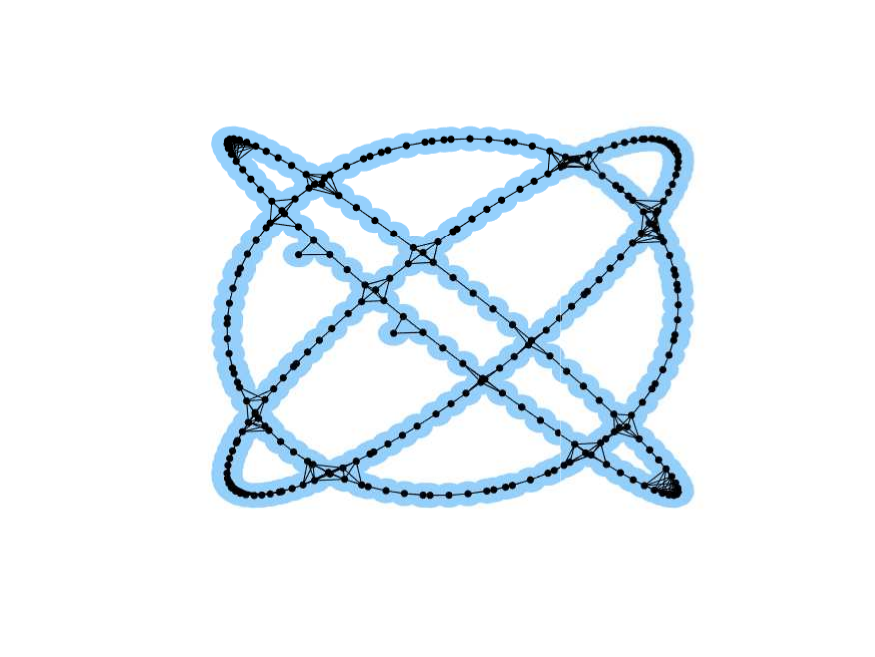}
				\label{rips1}}
			\quad
			\subfigure[ $\epsilon_2=0.12$, $\text{loops}=11$]{%
			\includegraphics[height=4.5cm,width=0.45\textwidth]{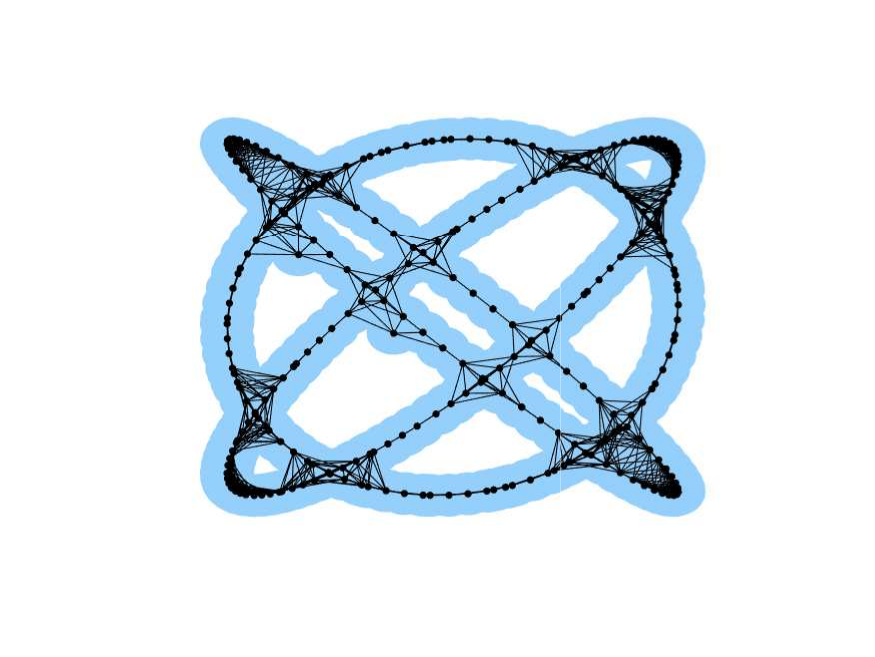}
				\label{rips2}}
			\quad
			\subfigure[ $\epsilon_3=0.2$, $\text{loops}=9$]{%
			\includegraphics[height=4.5cm,width=0.45\textwidth]{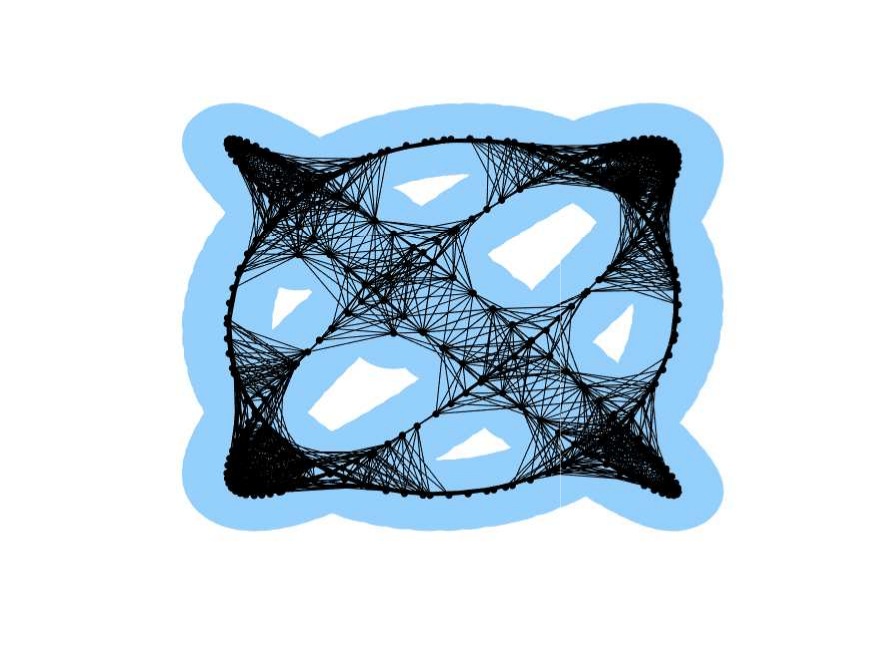}
    \label{rips3}}
    		\quad
			\subfigure[ $\epsilon_4=0.3$, $\text{loops}=2$]{%
			\includegraphics[height=4.5cm,width=0.45\textwidth]{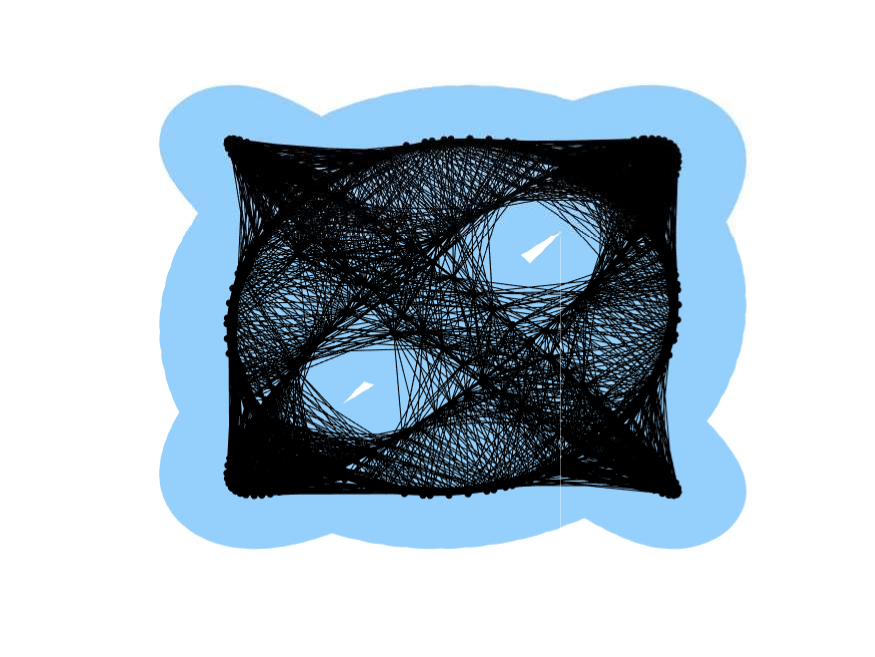}
    \label{rips4}}
		\end{center}
		\caption{ A diagram depicting the Rips filtration process. The point cloud in Figure \ref{fig:cloud} is transformed into a filtration of simplicial complexes using the Vietoris-Rips filtration, as illustrated in plots (a)-(d). This filtration involves drawing $\epsilon$ balls around each point in the point cloud and increasing $\epsilon$ gradually until all the balls merge.} \label{rips}
	\end{figure}

 \begin{figure}[!ht]
		\begin{center}
  			\subfigure{%
				\includegraphics[height=5cm,width=7cm]{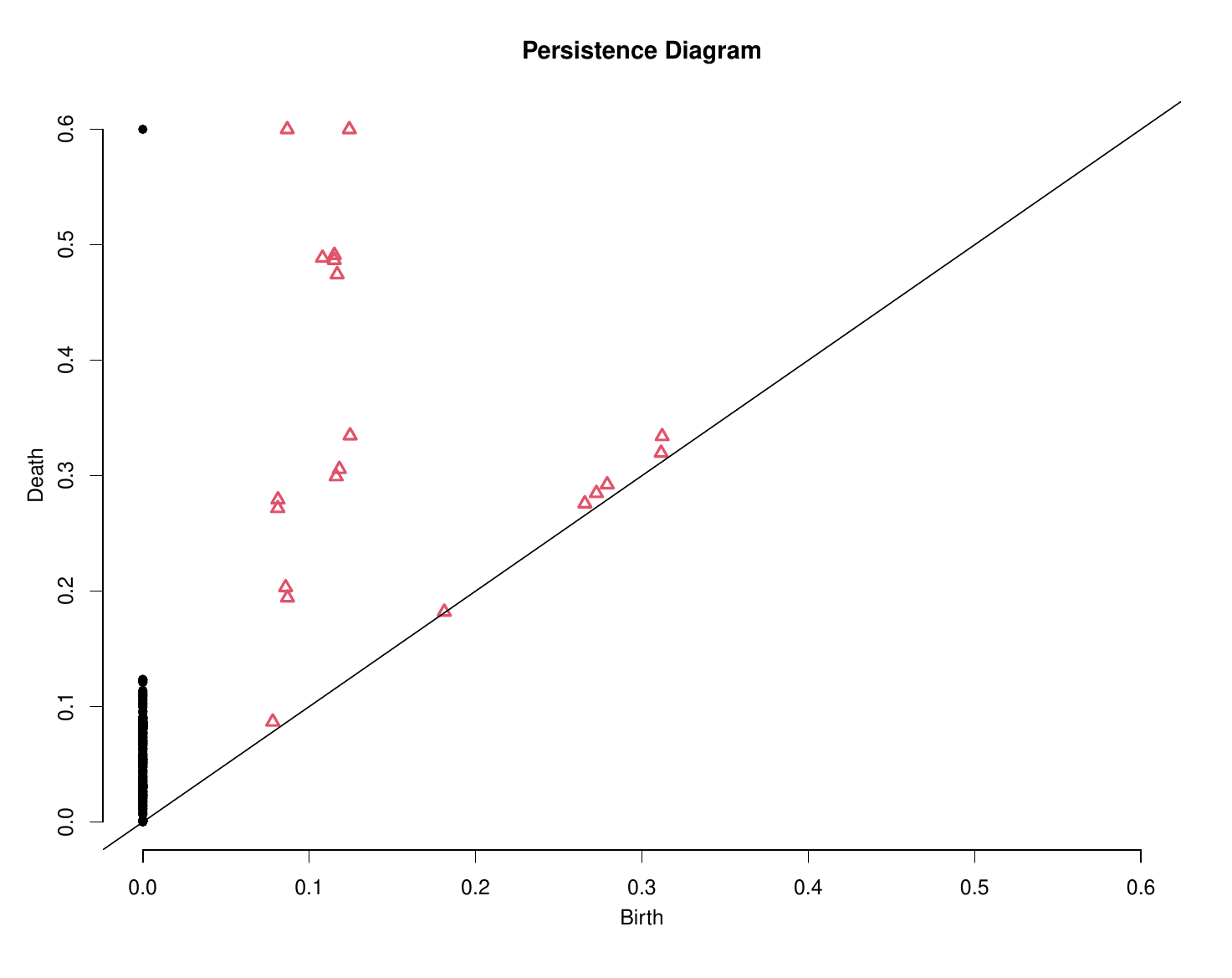}
				\label{perdiag}}
    \quad
			\subfigure{%
				\includegraphics[height=5cm,width=7cm]{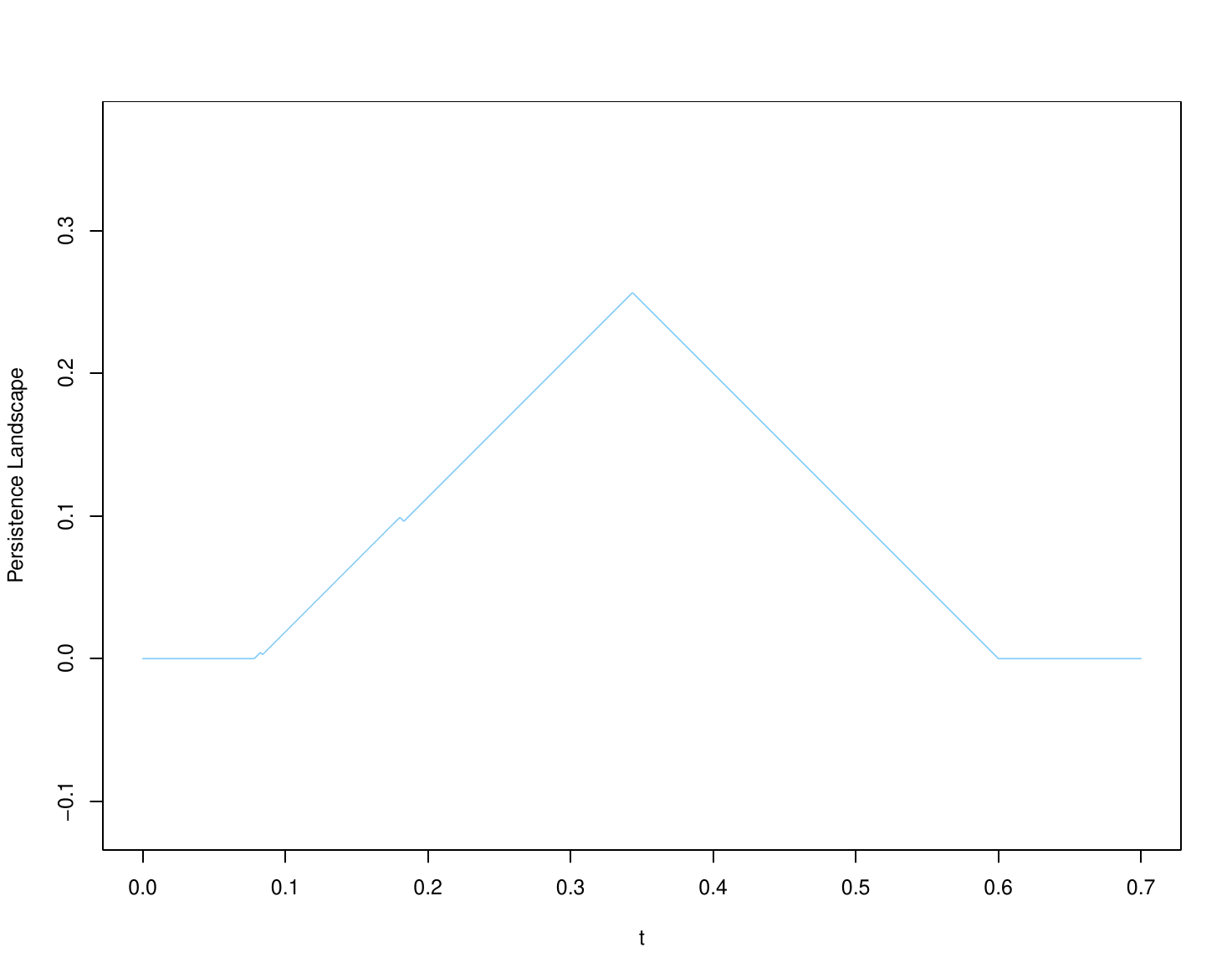}
				\label{fig:perlandscape}}
			\quad
   \end{center}
    \caption{Persistence diagram and the corresponding persistence landscape ($k=1$) for Figure \ref{rips}. The horizontal and vertical axes of the persistence diagram represent the length scales at which the features emerge, known as birth ($b$), and vanish, known as death ($d$), respectively. The points that are located far from the diagonal have longer lifetimes and are, therefore, considered more significant. For instance, the 13 off-diagonal red dots correspond to the birth and death of 13 significant holes present in the Rips Filtration. It is noteworthy that the creation of features precedes their destruction, resulting in no points lying below the diagonal. Moreover, at sufficiently large spatial scales, all points become connected to form a single point with an infinite lifetime, depicted as a black dot in the top-left corner of the persistence diagram.   }
    \label{pd_pl}
\end{figure}

\begin{figure*}[!ht]
    \centering
    \includegraphics[width=\textwidth]{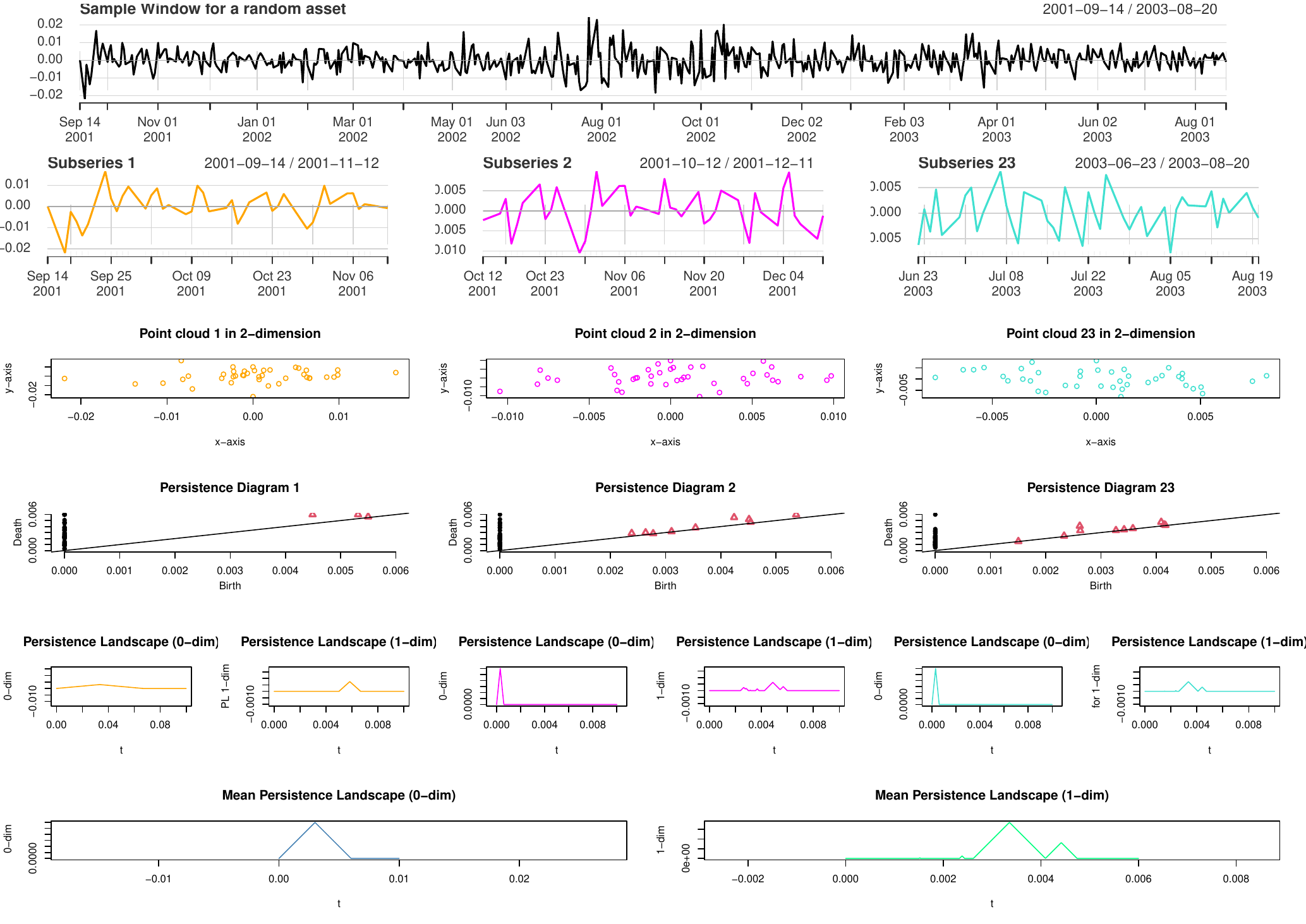}
    \caption{Graphical representation of the procedure utilizing a randomly selected asset in a sample window. The $L_1$ norm of the mean persistence landscape (0-dimensional) and mean persistence landscape (1-dimensional) serve as parameters $\alpha_i$ and $\beta_i$ for the asset within this time window.}
    \label{pd_pl2}
\end{figure*}

\clearpage

\subsection{In-sample analysis}\label{appendix:insample}

Table \ref{tab:in_sample} displays the in-sample performance metrics of the portfolios. This includes the average in-sample tracking error (objective function value) across various windows, as well as the mean values of other performance measures under consideration.

By analyzing the data from Tables \ref{tab:performance_analysis} and \ref{tab:in_sample}, it becomes evident that there is a decline in out-of-sample performance relative to in-sample performance. This is anticipated since the parameters are tailored for in-sample data. Specifically, for TE portfolios, the drop in performance is pronounced. This is logical, as all assets are aligned to minimize the tracking error based on in-sample data, leading to a potential overfitting scenario. Overfitting can be mitigated using penalty functions. However, traditional penalty methods like EN and SLOPE exhibit a more significant performance drop compared to TDA-based methods. This suggests that TDA effectively manages overfitting.

\begin{table*}[htbp]
\caption{\footnotesize{The in-sample performance for different sparse index tracking portfolios. Panel I reports results for data from October 11, 1999, to October 10, 2018 and Panel II from the Covid-19 January 1st, 2018 to March 8, 2023. The first column shows the results for the non-penalized $\text{TE}$ portfolio. Within both Panels, we report results for models that use both $\ell_1$ and $\ell_2$ penalty terms, i.e. Elastic-Net type penalty models, namely $\text{TDAEN11}$, $\text{TDAEN12}$, $\text{VolEN}$, and $\text{EN}$ (see Panels Ia and IIa). Moreover, results for models based on $\ell_1$ penalty term only, namely $\text{TDA} \ell_1$, $\text{Vol} \ell_1$, and $\text{SLOPE}$ with Lasso type penalty are reported (see Panels Ib and IIb). For a given measure, the best among all the values have been highlighted in \textbf{bold}.}} 
\scalebox{0.85}{%
\begin{tabular}{l|r|rrrr|rrr}
\hline
\multicolumn{9}{c}{Panel I:  Period 10/1999--10/2018}\\
\hline
&   & \multicolumn{4}{c|}{Panel Ia: $\ell_1$ and $\ell_2$ penalty} & \multicolumn{3}{c}{Panel Ib: $\ell_1$ penalty, 10/1999--10/2018} 
          \\
\hline
& \multicolumn{1}{l|}{TE} & \multicolumn{1}{l}{$\text{TDAEN11}$} & \multicolumn{1}{l}{$\text{TDAEN12}$} & \multicolumn{1}{l}{$\text{VolEN}$} & \multicolumn{1}{l|}{$\text{EN}$} & \multicolumn{1}{l}{$\text{TDA} \ell_1$} & \multicolumn{1}{l}{$\text{Vol} \ell_1$} & \multicolumn{1}{l}{$\text{SLOPE}$} \\
\hline
\textbf{Tracking performance:} & & & & & & & & \\
    $\text{ \quad TError}$ & \textbf{1.11E-04} & 4.50E-04 & 4.40E-04 & 4.54E-04 & 3.11E-04 & 4.54E-04 & 4.65E-04 & 6.77E-04 \\
    $\text{ \quad Correlation}$ & \textbf{99.97\%} & 99.52\% & 99.55\% & 99.54\% & 99.74\% & 99.51\% & 99.48\% & 98.64\% \\
    \textbf{Risk:} & & & & & & & & \\
     $\text{\quad Volatility}$&7.60E-02 & 7.37E-02 & \textbf{7.32E-02} & 7.59E-02 & 7.37E-02 & 7.37E-02 & 7.36E-02 & 7.56E-02\\
    $\text{ \quad DD}$    & \textbf{7.01E-05} & 2.91E-04 & 3.00E-04 & 4.62E-04 & 2.88E-04 & 2.81E-04 & 2.98E-04 & 2.01E-04 \\
    $\text{ \quad VaR}$   & 7.78E-03 & 7.48E-03 & 7.61E-03 & 7.78E-03 & \textbf{7.47E-03} & \textbf{7.47E-03} & 7.61E-03 & 7.70E-03 \\
    $\text{ \quad CVaR}$  & 1.12E-02 & \textbf{1.08E-02} & 1.09E-02 & 1.12E-02 & \textbf{1.08E-02} & \textbf{1.08E-02 }& 1.10E-02 & 1.11E-02 \\
    \textbf{Risk-adjusted performance:} & & & & & & & & \\
    $\text{ \quad SHR}$ & 3.01E-02 & 3.64E-02 & 3.48E-02 & 3.06E-02 & 3.64E-02 & \textbf{3.65E-02} & 3.63E-02 & 3.29E-02 \\
    $\text{ \quad IR}$ & 1.34E-01 & 8.69E-02 & 8.36E-02 & 3.02E-02 & 8.79E-02 & \textbf{9.01E-02} & 8.43E-02 & 8.65E-02 \\
\hline
\multicolumn{9}{c}{Panel II:  Period 1/2018--3/2023}\\
\hline
& &\multicolumn{4}{c|}{Panel IIa: $\ell_1$ and $\ell_2$ penalty} & \multicolumn{3}{c}{Panel IIb: $\ell_1$ penalty} \\
\hline
& \multicolumn{1}{l|}{TE} & \multicolumn{1}{l}{$\text{TDAEN11}$} & \multicolumn{1}{l}{$\text{TDAEN12}$} & \multicolumn{1}{l}{$\text{VolEN}$} & \multicolumn{1}{l|}{$\text{EN}$} & \multicolumn{1}{l}{$\text{TDA} \ell_1$} & \multicolumn{1}{l}{$\text{Vol} \ell_1$} & \multicolumn{1}{l}{$\text{SLOPE}$} \\
\hline
\textbf{Tracking performance:} & & & & & & & & \\
    $\text{ \quad TError}$ & \textbf{3.89E-08 }& 5.75E-04 & 5.52E-04 & 5.92E-04 & 8.05E-04 & 5.91E-04 & 6.04E-04 & 7.12E-04 \\
    $\text{ \quad Correlation}$ & \textbf{100.00\%} & 99.46\% & 99.47\% & 99.44\% & 98.62\% & 99.37\% & 99.35\% & 98.88\% \\
     \textbf{Risk:} & & & & & & & & \\
  $\text{\quad Volatility}$ & 5.6E-02 & 5.4E-02 & 5.39E-02 & \textbf{5.34E-02} & 5.49E-02 & 5.41E-02 & 5.36E-02 & 5.55E-02\\
    $\text{ \quad DD}$    & \textbf{2.72E-08} & 3.70E-04 & 3.54E-04 & 3.97E-04 & 5.49E-04 & 3.83E-04 & 4.03E-04 & 5.12E-04 \\
    $\text{ \quad VaR}$   & 9.30E-03 & 8.98E-03 & 8.98E-03 & 9.09E-03 & \textbf{8.96E-03} & 8.99E-03 & 9.09E-03 & 9.09E-03 \\
    $\text{ \quad CVaR}$  & 1.42E-02 & \textbf{1.36E-02} & \textbf{1.36E-02} & 1.37E-02 & 1.39E-02 & \textbf{1.36E-02} & 1.41E-02 & 1.39E-02 \\
 \textbf{Risk-adjusted performance:} & & & & & & & & \\
    $\text{ \quad SHR}$ & 3.611E-02 & 4.549E-02 & 4.516E-02 & 4.382E-02 & 4.524E-02 & \textbf{4.564E-02} & 4.411E-02 & 3.394E-02 \\
    $\text{ \quad IR}$ & 5.826E-02 & 7.578E-02 & \textbf{7.598E-02} & 7.498E-02 & 5.044E-02 & 7.574E-02 & 7.433E-02 & -1.030E-02 \\
    
\hline
\end{tabular}
}\label{tab:in_sample}
\end{table*}

 \end{document}